\documentclass{aastex62}
\usepackage{amsmath}

\received{April 24, 2019}
\revised{May 24, 2019}
\accepted{June 1, 2019}
\submitjournal{The Astrophysical Journal}
\shorttitle{KOI-3278: Einstein vs. Newton}
\shortauthors{Yahalomi et al.}

\begin{document}

\title{The Mass of the White Dwarf Companion in the Self-Lensing Binary KOI-3278: Einstein vs. Newton}

\correspondingauthor{Daniel A. Yahalomi}
\email{daniel.yahalomi@cfa.harvard.edu}

\author[0000-0003-4755-584X]{Daniel A. Yahalomi} 
\affiliation{Center for Astrophysics $\vert$ Harvard $\&$ Smithsonian,
60 Garden Street, Cambridge, MA 02138, USA}

\author[0000-0003-1525-5041]{Yossi Shvartzvald}
\affiliation{IPAC, Mail Code 100-22, Caltech, 1200 E. California Blvd., Pasadena, CA 91125, USA}

\author[0000-0002-0802-9145]{Eric Agol}
\affiliation{Department of Astronomy, Box 351580, University of Washington, Seattle, WA 98195-1580, USA}

\author[0000-0002-1836-3120]{Avi Shporer}
\affiliation{Department of Physics and Kavli Institute for Astrophysics
and Space Research, Massachusetts Institute of Technology,
Cambridge, MA 02139, USA}

\author[0000-0001-9911-7388]{David W. Latham}
\affiliation{Center for Astrophysics $\vert$ Harvard $\&$ Smithsonian,
60 Garden Street,
Cambridge, MA 02138, USA}

\author[0000-0002-0493-1342]{Ethan Kruse}
\affiliation{NASA Goddard Space Flight Center, 8800 Greenbelt Road,
Greenbelt, MD 20771, USA}

\author[0000-0002-9873-1471]{John M. Brewer}
\affiliation{Department of Astronomy, Yale University, 52 Hillhouse Avenue, New Haven, CT 06511, USA}
\affiliation{Department of Astronomy, Columbia University, 550 West 120th Street, New York, New York 10027}

\author[0000-0003-1605-5666]{Lars A. Buchhave}
\affiliation{DTU Space, National Space Institute, Technical University of Denmark, Elektrovej 328, DK-2800 Kgs. Lyngby, Denmark}

\author[0000-0003-3504-5316]{Benjamin J. Fulton}
\affiliation{California Institute of Technology, Pasadena, CA 91125, USA}
\affiliation{IPAC-NASA Exoplanet Science Institute, Pasadena, CA 91125
USA}

\author[0000-0001-8638-0320]{Andrew W. Howard}
\affiliation{California Institute of Technology, Pasadena, CA 91125, USA}

\author[0000-0002-0531-1073]{Howard Isaacson}
\affiliation{Department of Astronomy, University of California, Berkeley, CA 94720, USA}
\affiliation{University of Southern Queensland, Toowoomba, QLD 4350, Australia}

\author[0000-0003-0967-2893]{Erik A. Petigura}
\affiliation{California Institute of Technology, Pasadena, CA 91125, USA}
\affiliation{Hubble Fellow}

\author[0000-0002-8964-8377]{Samuel N. Quinn}
\affiliation{Center for Astrophysics $\vert$ Harvard $\&$ Smithsonian,
60 Garden Street,
Cambridge, MA 02138, USA}

\nocollaboration



\vspace{5mm}
\begin{abstract}

KOI-3278 is a self-lensing stellar binary consisting of a white-dwarf secondary orbiting a Sun-like primary star.  \cite{KruseAgol} noticed small periodic brightenings every 88.18 days in the Kepler photometry and interpreted these as the result of microlensing by a white dwarf with about 63$\%$ of the mass of the Sun.  We obtained two sets of spectra for the primary that allowed us to derive three sets of spectroscopic estimates for its effective temperature, surface gravity, and metallicity for the first time. We used these values to update the \citet{KruseAgol} Einsteinian microlensing model, resulting in a revised mass for the white dwarf of $0.539^{+0.022}_{-0.020} \, M\textsubscript{\(\odot\)}$. The spectra also allowed us to determine radial velocities and derive orbital solutions, with good agreement between the two independent data sets. An independent Newtonian dynamical MCMC model of the combined velocities yielded a mass for the white dwarf of $0.5122^{+0.0057}_{-0.0058} \, M\textsubscript{\(\odot\)}$. The nominal uncertainty for the Newtonian mass is about four times better than for the Einsteinian,  $\pm 1.1\%$ vs. $\pm 4.1\%$ and the difference between the two mass determinations is $5.2 \%$. We then present a joint Einsteinian microlensing and Newtonian radial velocity model for KOI-3278, which yielded a mass for the white dwarf of $0.5250^{+0.0082}_{-0.0089} \, M\textsubscript{\(\odot\)}$. This joint model does not rely on any white dwarf evolutionary models or assumptions on the white dwarf mass-radius relation. We discuss the benefits of a joint model of self-lensing binaries, and how future studies of these systems can provide insight into the mass-radius relation of white dwarfs.

\end{abstract}



\setlength{\tabcolsep}{3pt}
\section{Introduction} \label{sec:intro}
Stellar binaries with a compact companion (white dwarf, neutron star, or black hole), in which the compact object periodically lenses its primary as it passes in front of it, are called self-lensing stellar binaries. These systems were predicted as early as 1969 \citep{TrimbleThorne, LeibovitzHube, Maeder} and they provide an opportunity to test relativistic predictions, from the microlensed light curve, against dynamical predictions, from the spectroscopically observed radial velocities. Self-lensing systems also provide a window into the study of post-common envelope binaries, blue stragglers, and the formation of supernovae \citep{Preston, Zorotovic1, Zorotovic2, Kawahara}. \cite{KruseAgol} reported the discovery of the first such system: Kepler Object of Interest 3278.  KOI-3278 was initially classified as a transiting exoplanet candidate \citep{Kepler1, Kepler2} because the $Kepler$ light curve showed periodic dips that resembled the signal expected for a transiting planet. \cite{KruseAgol} noticed positive pulses with the same period as the transit-like dips, but offset in phase by close to half the period. They interpreted the dips as occultations of a white dwarf companion as it passed behind the Sun-like primary star and the pulses as magnifications due to gravitational microlensing by the white dwarf secondary as it passed in front of the primary.  \cite{KruseAgol} used the Kepler light curve to model the microlensing pulses as inverted transits. This approximation holds when the Einstein radius of the lens is small relative to the lensed source \citep{Agol}. Their model allowed them to derive a mass for the white dwarf relative to the mass of the primary.

Because spectroscopy of the primary star was not available, \cite{KruseAgol} were forced to rely on multiband photometry to estimate key stellar parameters for the primary.  They then used the Padova PARSEC stellar models \citep{PARSEC} to derive a mass for the primary.  Our follow-up spectroscopic observations provide improved estimates for the stellar parameters of the primary (effective temperature, surface gravity, and metallicity).  This allowed us to derive improved constraints for the mass and radius of the primary, again with the help of the same stellar models. We then reran essentially the same microlensing model as described in detail by \citet{KruseAgol}, but using our new stellar parameters. Thus the change compared to \citet{KruseAgol} in our value for the mass of the white dwarf companion stems mainly from our revision of the stellar parameters for the primary star. Our spectra also provide radial velocities suitable for a single-lined orbital solution, meaning the spectra of only the G star is seen. Together with our updated mass for the primary, this provides a dynamical mass for the white dwarf companion that depends only on Newtonian physics and the stellar models we adopted.  We thus have two independent predictions for the mass of the white dwarf companion: one from an Einsteinian microlensing model and one from a Newtonian dynamical model, both relying on the same stellar models. We then present a joint model for KOI-3278, that takes advantage of both Einsteinian and Newtonian models. Doing so allows us to independently solve for the mass and radius of the white dwarf using only isochrone fitting for the G star, dynamical equations to solve for the white dwarf mass, and microlensing equations to solve for the white dwarf radius.

In Section \ref{sec: spectroscopy}, we describe the methods used to determine stellar parameters and radial velocities from the spectroscopic observations. In Section \ref{sec: LC}, we then describe the MCMC model used to analyze the microlensing light curve and present the updated mass for the white dwarf based on Einstein's General Relativity. In Section \ref{sec: RV}, we present the MCMC model used to derive a single-lined spectroscopic orbital solution from the radial-velocity observations and present a dynamical mass for the white dwarf based on Newtonian physics. In Section \ref{sec: Joint} we present the joint MCMC model using both Einsteinian microlensing and Newtonian dynamical equations. In Section \ref{sec: discussion}, we compare the results from the two independent models and the joint model. We then discuss the implications of the joint model on the white-dwarf mass radius relation. Finally, we discuss future opportunities for studies of self-lensing binary systems. The code used for analysis is provided in a repository on GitHub.\footnote{https://github.com/dyahalomi/koi3278}


\pagebreak
\section{SPECTROSCOPIC OBSERVATIONS}
\label{sec: spectroscopy}

We mounted two independent campaigns to obtain suitable spectra, one with the the High Resolution Echelle Spectrometer \citep[HIRES,][]{HIRES} on the 10m Keck I telescope on Mauna Kea, HI, and the other with the Tillinghast Reflector Echelle Spectrograph \citep[TRES,][]{TRES} on the 1.5m Tillinghast Reflector at the Fred L. Whipple Observatory on Mt. Hopkins, AZ. Eight spectra of KOI-3278 were obtained with HIRES spread out over nearly four years between October, 2013 and April, 2017, supplemented by eight spectra obtained with TRES in the fall of 2017.  The HIRES spectra were obtained without use of the iodine gas-absorption cell. HIRES has higher resolving power than TRES, about 60,000 compared to 44,000, and not surprisingly those spectra have better SNR per resolution element than the TRES spectra, about 40 compared to 15, so we focused our efforts on the HIRES spectra for determining stellar parameters.  Three independent analyses were carried out, one using the Stellar Parameter Classification tool \citep[SPC;][]{Buchhave2012}, a second by John Brewer \citep[Brewer;][]{Brewer2016}, and a third using SpecMatch \citep[SpecMatch;][]{Petigura17b}.  SPC uses a correlation analysis of the observed spectra against a library of synthetic spectra calculated using Kurucz model atmospheres \citep{Kurucz} and does a multi-dimensional fit for the stellar parameters that give the highest peak correlation value.  The metallicity is assumed to have the same pattern of elemental abundances as the Sun. Brewer's analysis uses Spectroscopy Made Easy \citep[SME;][]{SME} to forward model the spectra to fit both the global stellar properties and individual abundances of 15 elements.  The method first fits $T_{eff}$, $\log{g}$, rotational broadening, and a scaled solar abundance pattern [M/H], allowing only calcium, silicon, and titanium abundances to vary independently.  The global parameters are then fixed while abundances of 15 elements are fit.  The whole procedure is then repeated, scaling this new abundance pattern rather than solar one in the first step.  Finally, a relation is used to fix the macroturbulence in order to solve for $\textit{v} \: \rm{sin} \: \textit{i}$.  The derived surface gravities are consistent with asteroseismically determined $\log{g}$ with an RMS scatter of 0.05 dex.  The relatively low S/N ($\sim$40) of the spectrum Brewer analyzed increases the uncertainties \citep{BrewerFischer} to $\sigma_{T_{\mathrm{eff}}} = 31$ K, $\sigma_{\log{g}} = 0.06$, and $\sigma_{[Fe/H]} = 0.02$. The SpecMatch algorithm is described in detail in \cite{Petigura17b} and \cite{Petigura15thesis}. In brief, SpecMatch fits five segments of HIRES spectrum by creating a synthetic spectrum by interpolating over a grid of model spectra computed by \cite{Coelho}. The $T_{eff}$, $\log{g}$, and [Fe/H], and $\textit{v} \: \rm{sin} \: \textit{i}$ of the synthetic spectrum are adjusted using a Non-Linear Least-Square optimizer \citep{Newville} until the best-matching spectrum is found.

Due to different model assumptions and calibrations between abundance analyses, the abundance uncertainties are only applicable in a relative sense within a single analysis technique. The results of these analyses are reported in Table \ref{tab: specPredictions}, along with the stellar parameters derived by the MCMC model in  \citet{KruseAgol}.  Note that the light from the white dwarf companion in the spectral regions used for the stellar parameter determinations has a negligible effect, so these parameters refer to the primary star.

\begin{table} [!htb]
\centering
\caption{Stellar Parameters for KOI-3278 from Spectroscopy}
\label{tab: specPredictions}
\begin{tabular}{ c  c  c  c  c}
\hline
Parameter & \cite{KruseAgol}  & SPC & Brewer & SpecMatch\\
\hline
$T_{\rm{eff}}$ (K) & $5568\pm39$  & $5435\pm50$ & $5384\pm45$ & $5490\pm60$ \\
$\log g$ (cgs) & $4.485\pm0.023$  & $4.59\pm0.10$ & $4.55\pm0.05$ & $4.62\pm0.07$ \\
$\rm{[Fe/H]}$ & $0.39\pm0.22$   & $0.22\pm0.08$ & $0.12\pm0.04$ & $0.16\pm0.04$ \\
$\textit{v} \: \rm{sin} \: \textit{i}$ ($\rm{km \: s^{-1}}$) & ...& $3.2\pm1.0$ & $3.6\pm1.0$ & $3.4\pm1.0$\\
\hline
\end{tabular}
\end{table}

\begin{table}[!htb]
\centering
\caption{Radial Velocity Observations}
\label{tab: specRV}
\begin{tabular}{cccc}
	\hline
	
	Spectrograph & Time (BJD) & RV ($\rm{km \: s^{-1}}$)  & RV Error ($\rm{km \: s^{-1}}$) \\
	
	\hline


    TRES  
      & 2458006.664944 &  -41.087 &  0.045\\
      & 2458009.684164 &  -43.562 &  0.064\\
      & 2458019.772179 &  -46.711 &  0.059\\
      & 2458038.615663 &  -29.351 &  0.079\\
      & 2458052.616284 &  -11.535 &  0.056\\
      & 2458063.641604 &   -8.130 &  0.063\\
      & 2458070.641157 &  -11.215 &  0.080\\
      & 2458081.601247 &  -24.260 &  0.070\\
	  & & & \\
	
	HIRES  
	  & 2456585.763935 & -28.888 & 0.089\\
      & 2456909.848497 & -9.044 & 0.086\\
	  & 2457579.984325 & -46.575 & 0.118\\
	  & 2457581.005670 & -46.524 & 0.139\\
	  & 2457652.901655 & -40.145 & 0.133\\
	  & 2457703.779060 & -8.813 & 0.072\\
	  & 2457829.106551 & -39.762 & 0.168\\
	  & 2457853.094255 & -40.780 & 0.149\\
	
	\hline

\end{tabular}
\end{table}

The same spectra that were used to derive stellar parameters for KOI-3278 were also used to derive radial velocities.  Telluric lines in the A and B bands of oxygen were used to establish the zero point for the HIRES velocities as documented by \citet{Nidever2002} and \citet{Chubak2012}.  The TRES velocities were derived using a correlation analysis that adopted a template constructed by co-adding the observed spectra after shifting to a common wavelength scale.  The resulting relative velocities were then shifted to the IAU System using run-to-run offsets (stable to better than $0.015 \, \rm{km \: s^{-1}}$) based on nightly observations of standard stars. The radial velocities are reported in Table \ref{tab: specRV}.

\begin{figure}[!htb]
\centering
\includegraphics[width=.77\textwidth]{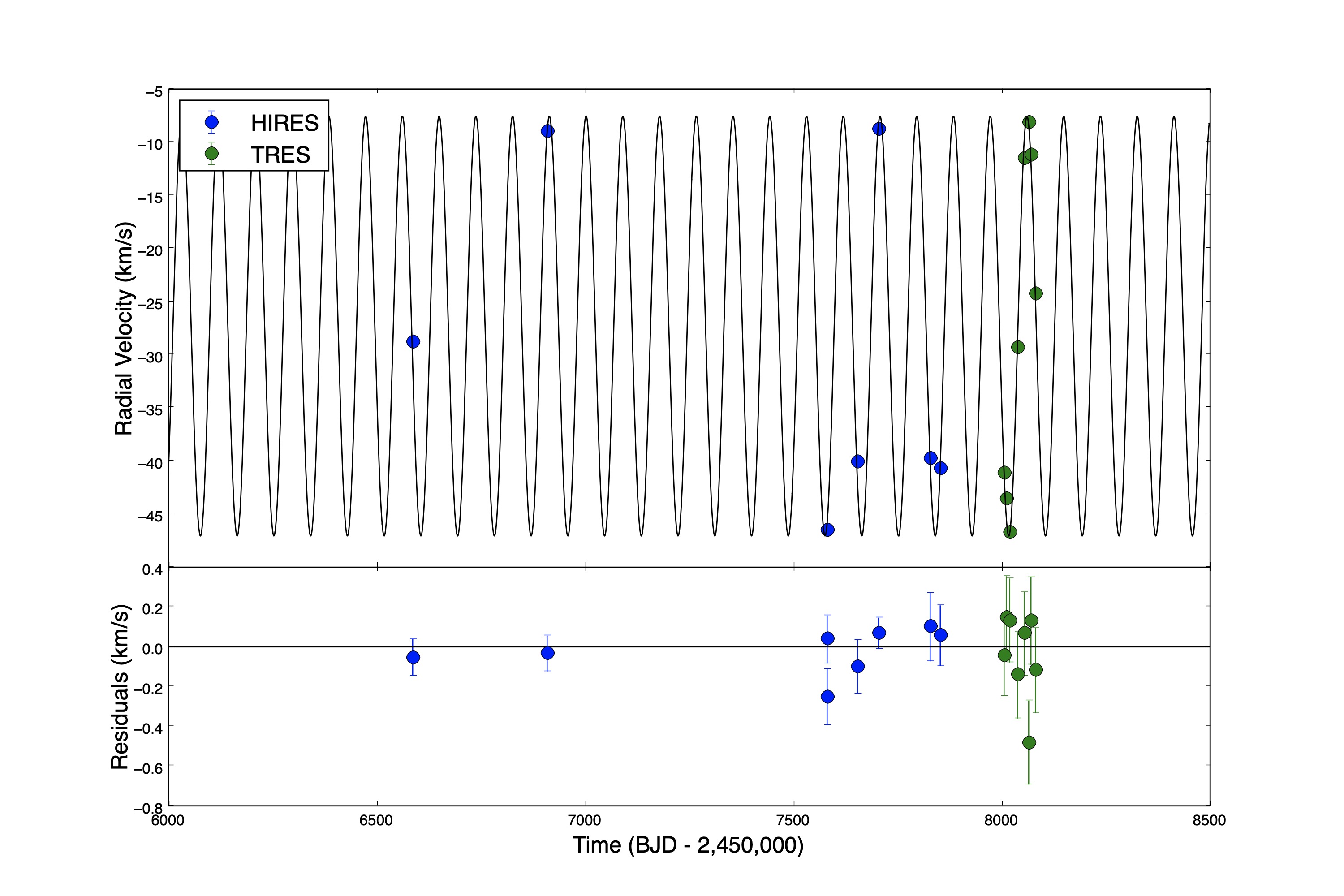}
\caption{Radial velocity observations from TRES and HIRES over time. The periodic curve shows the maximum-likelihood model from the MCMC fit to both TRES and HIRES radial velocity observations. The uncertainties in radial velocity observations are the reported errors from the spectroscopic radial velocities added in quadrature with the MCMC modelled radial velocity jitter terms.}
\label{fig: RVplot}
\end{figure}

The HIRES observations cover 18 orbital cycles, so they provide a much stronger constraint on the orbital period. However, six of the HIRES velocities were obtained as close pairs during individual observing runs, so effectively only five epochs are represented, and only one epoch lands in the second half of the orbital phase.  The TRES observations on the other hand are well distributed across the orbital phase, including velocities near both $\gamma$ crossings.  Thus, the orbital eccentricity is better constrained by the TRES observations. The complementary nature of the TRES and HIRES radial velocity observations can be clearly seen in Figure \ref{fig: RVplot}.

\pagebreak
\section{Updated Einsteinian Microlensing Model} \label{sec: LC}
We updated the microlensing model for KOI-3278 with a modified code that follows very closely the procedure used by \citet{KruseAgol}.  The updated Einsteinian microlensing model does not use any of the radial velocity data from the spectroscopic observations. With spectroscopic constraints on the isochrones, we no longer fit the apparent magnitude of the system. This was no longer necessary as our spectroscopic constraints allow us to constrain the SED. We also were able to remove all assumptions on the dust distribution. In order to fit for the white dwarf parameters, without radial velocity information, we were forced to assume a mass-radius relationship of the white dwarf. Similarly to \citet{KruseAgol}, we adopted the Nauenberg relation for the zero-temperature white dwarf, which can be seen in Equation \ref{eq: Nauenberg} - where $M_{Ch} = 1.454 \, M\textsubscript{\(\odot\)}$ is the Chandrasekhar mass \citep{Nauenberg}:

\begin{equation} \label{eq: Nauenberg}
    R_2 = 0.0108 \, \Bigg[ {\Big(\frac{M_2}{M_{Ch}}\Big)^{-\frac{2}{3}} - \Big(\frac{M_2}{M_{Ch}}\Big)^\frac{2}{3}} \Bigg] ^{\frac{1}{2}}.
\end{equation}

In order to model parameters for the system, we use \texttt{emcee}: a python implementation of Goodman and Weare's affine invariant ensemble sampler for Markov chain Monte Carlo (\citealp[Goodman and Weare;][]{GoodmanWear}, \citealp[\texttt{emcee};][]{emcee}). We used the same values from the Kepler time series photometry of time, flux, and flux error as used in \cite{KruseAgol}, and that can be accessed from their public GitHub.\footnote{https://github.com/ethankruse/koi3278} As was found in \citet{KruseAgol}, the initial MCMC fit produced a reduced chi-square value slightly larger than unity. By inflating the Kepler time series reported errors by a factor of 1.13, the MCMC microlensing fit returned a reduced chi-square that approached unity. As the Kepler time series photometry has relatively uniform errors, this functions similarly to fitting for a photometric jitter term in the MCMC model and allows our MCMC modeling errors to be consistent with those used in \citet{KruseAgol}.

We modeled the light curve following the \cite{KruseAgol} code with slight modifications. We made the following two changes: (1) we removed constraints on the Padova PARSEC isochrone models based on available photometry for KOI-3278 and replaced them Gaussian priors around the spectroscopic estimates of $T_{\rm{eff}}$, $\log g$, and [Fe/H]; (2) we removed several modeling parameters, namely: distance, systematic magnitude errors, dust scale height, and the total extinction and the corresponding assumptions that were required to model these parameters.

In the microlensing model, we have 10 fitted parameters: period, transit time, eccentricity ($e$) and longitude of periapsis of the G star ($\omega$) as $e \cos \omega$ and $e \sin \omega$, impact parameter ($b$), progenitor white dwarf mass, current white dwarf mass, current G star mass, metallicity, and log age of the system. We model $e \cos \omega$ and $e \sin \omega$ instead of e and $\omega$ because it increases convergence speed. Due to this, we must apply a prior of 1/$e$ at each step of the model \citep{Exofast, KruseAgol}. The modelling parameters and priors are listed in Table \ref{tab: compare_lc_spec_results}.

The progenitor white dwarf mass is the initial mass of the white dwarf. \citet{KruseAgol} found that in most cases, the white dwarf initial and final masses fall within $10\%$ of the Kalirai initial-final white dwarf mass relation \citep{Kalirai}. Therefore, we place a Gaussian prior for $M_2$ to fall with $10\%$ of the mass prediction from the Kalirai prediction based on $M_{2, init}$. If we define the Kalirai white dwarf mass prediction as $M_{\rm{2, p}} = 0.109 \, M_{2,init} + 0.394$, we add a chi-square penalty of $\chi^2: (M_2 - M_{\rm{2, p}})^2 / (0.1 \, M_{\rm{2, p}})^2 $ at each step. The limb darkening parameters were modeled based on a fit to stellar atmosphere predictions for the quadratic limb-darkening coefficients as a function effective temperature, surface gravity, and metallicity from \citep{Sing}, as done in \citet{KruseAgol}, resulting in the relations

\begin{equation}
\begin{gathered} 
\label{eq: limb darkenning}
    u_1 = 0.4466 - 0.196 \Big(\frac{T_{eff, 1}}{10^3} - 5.5 \Big) + 0.00692 \log_{10} \Big(\frac{g_1}{10^{4.5}} \Big) + 0.0865 [Fe/H]_1 
    \\
    u_2 = 0.2278 - 0.128 \Big(\frac{T_{eff, 1}}{10^3} - 5.5 \Big) - 0.00458 \log_{10} \Big(\frac{g_1}{10^{4.5}} \Big) - 0.0506 [Fe/H]_1.
\end{gathered}
\end{equation}

\noindent Finally, in order to model the evolution of the white dwarf, we used the cooling models computed by Bergeron and collaborators \citep{HolbergBergeron, KowalskiSaumon, Tremblay, Bergeron} which may be found on their website.\footnote{http://www.astro.umontreal.ca/$\sim$bergeron/CoolingModels/}

\begin{figure}[!htb]
\centering
\includegraphics[width=.9\textwidth]{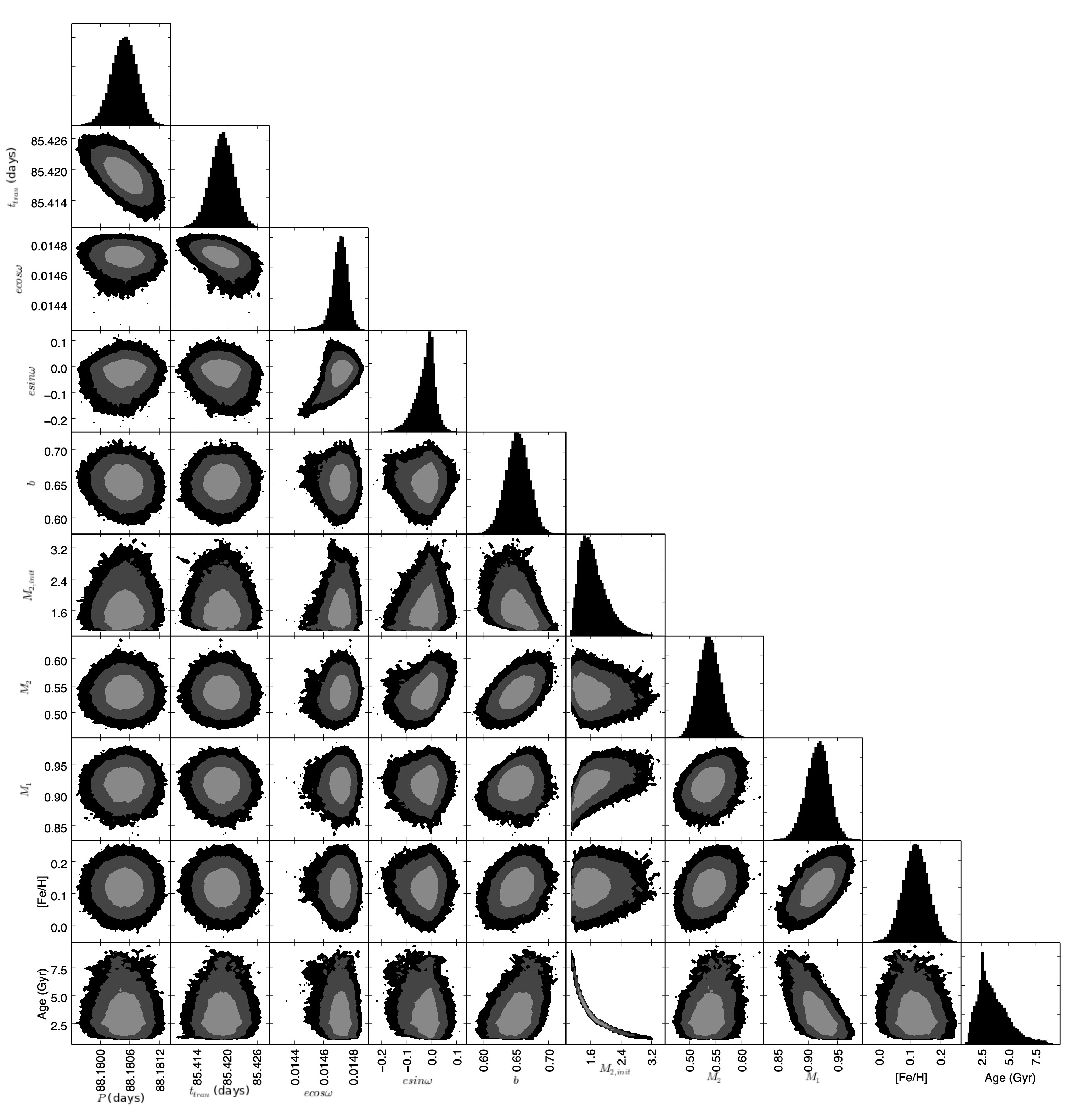}
\caption{Contour plots showing the $1 \sigma$, $2 \sigma$, and $3 \sigma$ constraints on pairs of parameters for the updated microlensing model using the Brewer's stellar estimates and Kepler photometry. Masses are all in units of solar masses.}
\vspace{2mm}
\label{fig: lc_spec_triangle}
\end{figure}

We ran three MCMC models with 100,000 steps and 50 walkers, and we discarded the first 20,000 steps as burn in. We ran an independent MCMC model for each of the SPC, Brewer, and SpecMatch estimates of G star parameters from the HIRES spectroscopy. We tested for convergence by enforcing that the number of independent draws was greater than 1,000 and determining the Gelman-Rubin statistic for each modeled parameter \citep{Ford, RadVel}. The maximum Gelman-Rubin value for the chains is 1.17, which was for age in the MCMC model using SPC's stellar parameter estimates. All other modeled parameters had a Gelman-Rubin statistic less than 1.1. The median modeled MCMC parameters have a reduced chi-square of $\frac{\chi^2}{\rm{DOF}}_{SPC} = 1.03$, $\frac{\chi^2}{\rm{DOF}}_{Brewer} = 1.03$, and  $\frac{\chi^2}{\rm{DOF}}_{SpecMatch} = 1.03$, respectively. The corner plot for the MCMC model with Brewer's estimates of stellar parameters can be seen in Figure \ref{fig: lc_spec_triangle}, and the predicted parameters for all three microlensing models can be seen compared to the original \cite{KruseAgol} model in Table \ref{tab: compare_lc_spec_results}. Using Brewer's stellar estimates as priors on MCMC model, our updated microlensing value for the white dwarf mass is $0.539^{+0.022}_{-0.020} \, M\textsubscript{\(\odot\)}$. For a longer discussion regarding the  mass estimate of the white dwarf, see Section \ref{sec: discussion}.

\begin{table} 
\centering
\caption{Parameters from the Microlensing Model}
\label{tab: compare_lc_spec_results}
\begin{tabular}{ c  c  c  c  c  c }
\bigskip\\

\multicolumn{6}{c}{G Star}\\
\hline
Parameter & Prior\footnote{\label{fn: priorLC} Priors adopted in the MCMC model. If no prior is listed, then the parameter is a derived parameter. $\mathcal{U}(x,y)$ denotes a uniform distribution between x and y. $\mathcal{N}(\mu, \sigma^2$) denotes a Gaussian distribution centered at $\mu$ with width of $\sigma$.} & \cite{KruseAgol} & SPC & Brewer & SpecMatch \\
\hline 

$M_1$ ($M\textsubscript{\(\odot\)}$)  & $\mathcal{U}$(0, $\infty$) & $1.042^{+0.028}_{-0.058}$ & $0.959^{+0.027}_{-0.029}$ & $0.918^{+0.018}_{-0.020}$  & $0.959^{+0.021}_{-0.021}$ \\

$R_1$ ($R\textsubscript{\(\odot\)}$) & ... &$0.964^{+0.034}_{-0.054}$ & $0.893^{+0.025}_{-0.024}$ & $0.857^{+0.018}_{-0.017}$ & $0.890^{+0.021}_{-0.020}$  \\

[Fe/H]$_1$ & $\mathcal{N}(\mu_{\rm{spec}}, \sigma^2_{\rm{spec}})$\footnote{\label{fn: specPrior} Gaussian prior around the spectroscopic estimates of the stellar parameters (effective temperature, metallicity, and surface gravity, respectively) from SPC, Brewer, and SpecMatch. Metallicity is a free parameter in the model. Effective temperature and surface gravity are derived parameters in the model. See Table \ref{tab: specPredictions} for spectroscopic estimates ($\mu_{\rm{spec}}$) and errors ($\sigma_{\rm{spec}}$).} & $0.39^{+0.22}_{-0.22}$  & $0.222^{+0.080}_{-0.076}$ &  $0.122^{+0.040}_{-0.040}$ & $0.160^{+0.040}_{-0.039}$  \\

Age$_1$ (Gyr) & $\mathcal{N}(0.89, 0.15^2)$\footnote{If age of the star is less than the spin-down age of star, Gaussian prior around spin-down age = $0.89 \pm 0.15$ Gyr \citep{KruseAgol}.} & $1.62^{+0.93}_{-0.55}$ & $2.74^{+1.66}_{-0.92}$ &  $3.3^{+1.7}_{-1.0}$ & $2.57^{+1.26}_{-0.83}$ \\

$T_{\rm{eff,1}}$ (K) & $\mathcal{N}(\mu_{\rm{spec}}, \sigma^2_{\rm{spec}})^{\rm{\ref{fn: specPrior}}}$ & $5568^{+40}_{-39}$ & $5441^{+50}_{-51}$ & $5391^{+43}_{-44}$ & $5494^{+58}_{-60}$ \\

$\log g_1$ & $\mathcal{N}(\mu_{\rm{spec}}, \sigma^2_{\rm{spec}})^{\rm{\ref{fn: specPrior}}}$ &$4.485^{+0.026}_{-0.020}$ &  $4.516^{+0.016}_{-0.019}$ & $4.533^{+0.013}_{-0.018}$ & $4.519^{+0.014}_{-0.016}$ \\
\hline

\bigskip\\
\multicolumn{6}{c}{White Dwarf}\\
\hline 
Parameter & Prior$^{\rm{\ref{fn: priorLC}}}$ & \cite{KruseAgol} & SPC & Brewer & SpecMatch \\
\hline 

$M_2$ ($M\textsubscript{\(\odot\)}$)  & $\mathcal{N}$($M_{\rm{2, p}}$, $[0.1 M_{\rm{2, p}}]^2$)\footnote{Gaussian prior for $M_2$ to fall within $10\%$ of the Kalirai relation \citep{Kalirai}. See discussion in Section \ref{sec: LC}. } & $0.634^{+0.047}_{-0.055}$ & $0.568^{+0.028}_{-0.027}$  & $0.539^{+0.022}_{-0.020}$ &  $0.567^{+0.026}_{-0.025}$ \\

$M_{\rm{2,init}}$ ($M\textsubscript{\(\odot\)}$) & $\mathcal{U}$(0, $\infty$) & $2.40^{+0.70}_{-0.53}$ & $1.83^{+0.53}_{-0.37}$  & $1.64^{+0.43}_{-0.28}$ & $1.89^{+0.50}_{-0.35}$ \\

$R_2$ ($R\textsubscript{\(\odot\)}$) & ... & $0.01166^{+0.00069}_{-0.00056}$  & $0.01249^{+0.00036}_{-0.00037}$ & $0.01288^{+0.00029}_{-0.00029}$ & $0.01250^{+0.00033}_{-0.00034}$ \\

$R_E$ ($R\textsubscript{\(\odot\)}$) & ... & $0.02305^{+0.00094}_{-0.00107}$  & $0.02171^{+0.00065}_{-0.00057}$ & $0.02101^{+0.00048}_{-0.00045}$   & $0.02167^{+0.00056}_{-0.00052}$ \\

MS Age (Gyr)  & ... & $0.96^{+0.90}_{-0.53}$ & $1.94^{+1.67}_{-0.92}$  & $2.5^{+1.7}_{-1.0}$   & $1.79^{+1.26}_{-0.83}$ \\

Cooling Age (Gyr) & ... & $0.663^{+0.065}_{-0.057}$  & $0.794^{+0.051}_{-0.051}$  & $0.861^{+0.048}_{-0.046}$ & $0.771^{+0.054}_{-0.051}$\\

$T_{\rm{eff,2}}$ (K) & ... & $9960^{+700}_{-760}$  & $8770^{+390}_{-350}$  & $8280^{+280}_{-260}$ & $8860^{+380}_{-350}$ \\

$L_{WD} (L_\odot)$  & ... & $0.00120^{+0.00024}_{-0.00022}$  & $0.000828^{+0.000111}_{-0.000093}$  & $0.000700^{+0.000075}_{-0.000065}$ &  $0.000864^{+0.000118}_{-0.000099}$ \\
\hline

\bigskip\\
\multicolumn{6}{c}{Binary System}\\
\hline
Parameter & Prior$^{\rm{\ref{fn: priorLC}}}$ & \cite{KruseAgol} & SPC & Brewer & SpecMatch \\
\hline 

$P$ (days) &  $\mathcal{U}$(0, $\infty$)  & $88.18052^{+0.00025}_{-0.00027}$ & $88.18051^{+0.00025}_{-0.00027}$   & $88.18050^{+0.00025}_{-0.00026}$ & $88.18051^{+0.00025}_{-0.00026}$ \\

$t_{\rm{tran}}$ (BJD $- 2,455,000$) &   $\mathcal{U}$(-$\infty$, $\infty$) &  $85.4190^{+0.0023}_{-0.0023}$  & $85.4191^{+0.0023}_{-0.0024}$ & $85.4192^{+0.0023}_{-0.0024}$  & $85.4190^{+0.0023}_{-0.0023}$ \\

$e\: \rm{cos} \: \omega$ &  $\mathcal{U}$(-1, 1) & $0.014713^{+0.000047}_{-0.000062}$  &  $0.014715^{+0.000046}_{-0.000054}$ & $0.014718^{+0.000045}_{-0.000051}$ & $0.014717^{+0.000044}_{-0.000052}$ \\

$e\: \rm{sin} \: \omega$  &  $\mathcal{U}$(-1, 1) & $0.000^{+0.049}_{-0.054}$ & $-0.013^{+0.030}_{-0.056}$ &   $-0.017^{+0.027}_{-0.045}$ & $-0.010^{+0.031}_{-0.051}$ \\

$b$ & $\mathcal{U}$(-$\infty$, $\infty$) & $0.706^{+0.020}_{-0.026}$  &  $0.674^{+0.018}_{-0.018}$  & $0.653^{+0.017}_{-0.018}$  & $0.673^{+0.016}_{-0.016}$ \\

$e$ & ... & $0.032^{+0.056}_{-0.016}$ & $0.029^{+0.046}_{-0.013}$ &  $0.028^{+0.037}_{-0.012}$ & $0.028^{+0.041}_{-0.012}$  \\

$\omega$ (deg) & ... & $2^{+72}_{-76}$ & $-41^{+91}_{-37}$  &  $-48^{+83}_{-28}$ &  $-35^{+89}_{-41}$ \\

$a$ (AU) & ... & $0.4605^{+0.0064}_{-0.0103}$  & $0.4464^{+0.0047}_{-0.0047}$ &  $0.4394^{+0.0032}_{-0.0033}$ & $0.4463^{+0.0040}_{-0.0039}$  \\

$a/R_1$ & ... & $102.8^{+3.7}_{-2.4}$ &  $107.6^{+2.1}_{-2.1}$ &  $110.4^{+1.7}_{-1.9}$   & $107.9^{+1.8}_{-1.9}$ \\

$i$ (deg) & ... & $89.607^{+0.027}_{-0.020}$  & $89.641^{+0.016}_{-0.016}$ & $89.661^{+0.013}_{-0.014}$ & $89.643^{+0.014}_{-0.014}$ \\

$K_1$ ($\rm{km \: s^{-1}}$)  & ... & $21.53^{+0.97}_{-0.98}$ & $20.51^{+0.65}_{-0.60}$ & $20.07^{+0.58}_{-0.54}$   & $20.49^{+0.62}_{-0.59}$ \\

$u_1$ & ... & ...& $0.478^{+0.012}_{-0.012}$   & $0.4787^{+0.0095}_{-0.0092}$ & $0.462^{+0.012}_{-0.012}$ \\

$u_2$  & ... & ...& $0.2241^{+0.0077}_{-0.0075}$ & $0.2356^{+0.0059}_{-0.0059}$ & $0.2204^{+0.0079}_{-0.0075}$ \\

Magnification - 1 & ... &  $0.001003^{+0.000053}_{-0.000049}$  & $0.000988^{+0.000046}_{-0.000042}$ &  $0.000977^{+0.000042}_{-0.000040}$  & $0.000989^{+0.000044}_{-0.000042}$ \\

$F_2/F_1$ & ... & $0.001125^{+0.000039}_{-0.000039}$  & $0.001127^{+0.000039}_{-0.000039}$ & $0.001127^{+0.000039}_{-0.000039}$ &  $0.001127^{+0.000039}_{-0.000039}$ \\

\hline
\end{tabular}
\end{table}


\clearpage
\section{Newtonian Dynamical Model} \label{sec: RV}
As described in Section \ref{sec: spectroscopy}, the HIRES and TRES spectra also provide single-lined radial velocities for the Sun-like primary star in the KOI-3278 binary system.  The velocities are reported in Table \ref{tab: specRV}. Again using MCMC modelling, we derived orbital solutions, both for the individual velocity sets and for the combined velocities with the offset between the two velocity sets allowed to be a free parameter. This Newtonian dynamical model does not use any photometry in order to constrain the parameters of the stellar binary.

\subsection{Keplerian Solver} \label{subsec: Keplerian Solver}
We solve the Keplerian problem for the radial velocity of the host star as a function of time, using the following equations:

\begin{equation}
	\tau = t_{\rm{tran}} + \sqrt{1 - e^2} \frac{P}{2 \pi} \; \Big[\frac{e \sin (\frac{\pi}{2} - \omega)}{1 + e \cos (\frac{\pi}{2} - \omega) } \; - \frac{2}{\sqrt{1 - e^2}} \; \tan^{-1} \big(\frac{\sqrt{1 - e^2} \tan (\frac{\pi}{4} - \frac{\omega}{2})}{1+e} \big) \Big],
\end{equation}

\begin{equation}
	M = [\frac{2 \pi}{P} * (t-\tau)] \pmod{2 \pi},
\end{equation}

\begin{equation} \label{eq: keplerEq}
	M = E - e \sin E,
\end{equation}

\begin{equation}
	\nu = 2 \tan^{-1} \Big [\sqrt{\frac{1+e}{1-e}} \tan (\frac{E}{2}) \Big]  \pmod{2 \pi},
\end{equation}

\begin{equation}
	RV = K \Big[ e \cos \omega + \cos (\nu + \omega) \Big] + \gamma.
\end{equation}

Solving this set of equations based on the modeling parameters gives a predicted RV as a function of time. In order to solve Kepler's Equation \ref{eq: keplerEq} we implement an iterative solution via Newton's method \citep{Zechmeister}.

\subsection{MCMC Model} \label{subsec: RV Model}
Once we have a model for radial velocity as a function of time from our Keplerian solver, we then compare this modeled RV with the observed RVs as the MCMC maximizes Equation \ref{eq: likelihoodFuncRV} \citep{Christiansen},

\begin{equation} \label{eq: likelihoodFuncRV}
	\ln \mathcal{L}_{RV} = -\sum_{i} \Big[ \frac{[v_i - v_m(t_i)] ^ 2}{2 (\sigma_i ^2 + \sigma_j^2)} + \ln \sqrt{2 \pi (\sigma_i ^2 + \sigma_j^2})  \Big],
\end{equation}
\noindent where $v_i$ is the observed velocity, $v_m(t_i)$ is the modeled velocity at time $t_i$, $\sigma_i$ is the reported error, and $\sigma_j$ is the ``velocity jitter'' term needed to achieve a reduced $\chi^2$ that approaches unity for the velocity residuals when added in quadrature.

We used this MCMC model to derive orbital solutions, initially for each of the two independent sets of velocities, with 7 free parameters: period ($P$), transit time ($t_{\rm{tran}}$), eccentricity ($e$) and longitude of periapsis of the G star ($\omega$) as $e \cos \omega$ and $e \sin \omega$, radial velocity semi-amplitude ($K$), center of mass velocity ($\gamma$), and stellar jitter ($\sigma_j$). 

The orbital parameters and priors used in the modelling for the individual TRES and HIRES models are reported in Table \ref{tab: RVfits}. Remarkably, the two independent orbital solutions yield a semi-amplitude, $K$, that differ by only $0.6\%$. From the parameter $K$, one can estimate the mass ratio between the white dwarf secondary and the G star primary.

We ran an MCMC model with 100,000 steps and 100 walkers for both independent sets of TRES and HIRES spectroscopy, and we threw out the first 2,000 steps as burn in. We tested for convergence by enforcing that the number of independent draws was greater than 1,000 and determining the Gelman-Rubin statistic for each modeled parameter \citep{Ford, RadVel}. The maximum Gelman-Rubin value of the chains is 1.002. The median modeled MCMC parameters have a chi-square of $\frac{\chi^2}{\rm{DOF}}_{HIRES} = 1.14$ and $\frac{\chi^2}{\rm{DOF}}_{TRES} = 1.14$, respectively. The error used in determining the reduced chi-square statistic is  the reported errors added in quadrature to the MCMC modelled radial velocity jitter.

As mentioned previously, and can be seen in Figure \ref{fig: RVplot}, the two sets of velocity observations complement each other rather well, and we modeled the combined velocities with one additional parameter for the offset between the zero points of the two velocity sets, $\gamma_o$. The orbital parameters for the combined solution are reported in Table \ref{tab: RVfits}. The absolute $\gamma$ velocities for the two solutions agree quite well, differing by only 0.03 $\rm{km \: s^{-1}}$.  This value is typical for the uncertainty in establishing the zero point for velocities of Sun-like stars on an absolute system.  In general, the improvement in the errors estimated for the orbital parameters from the combined solution is quite impressive.

\begin{table}
\centering
\caption{Orbital Predictions from MCMC Model with Different Spectroscopic Constraints}
\label{tab: RVfits}
\begin{tabular}{ c  c  c  c  c }

\hline

Parameter & Prior\footnote{\label{fn: priorLC} Priors adopted in the MCMC model. If no prior is listed, then the parameter is a derived parameter. $\mathcal{U}(x,y)$ denotes a uniform distribution between x and y. $\mathcal{N}(\mu, \sigma^2$) denotes a Gaussian distribution centered at $\mu$ with width of $\sigma$.} & TRES & HIRES & TRES and HIRES\\

\hline

$P$ (days) & $\mathcal{U}$(0, $\infty$)  & $88.38^{+1.20}_{-0.84}$ & $88.171^{+0.047}_{-0.129}$ & $88.189^{+0.014}_{-0.014}$ \\

$t_{\rm{tran}}$ (BJD $- \, 2,450,000$) & $\mathcal{U}$(-$\infty$, $\infty$)  & $4991^{+29}_{-42}$  & $4997.5^{+2.3}_{-1.0}$ & $4997.21^{+0.37}_{-0.37}$ \\

$e \cos\omega$ &  $\mathcal{U}$(-1, 1)  & $0.0080^{+0.0132}_{-0.0063}$  & $0.0098^{+0.0195}_{-0.0077}$  & $0.0045^{+0.0055}_{-0.0034}$ \\

$e \sin\omega$ & $\mathcal{U}$(-1, 1)  & $-0.0050^{+0.0100}_{-0.0197}$ & $-0.011^{+0.022}_{-0.080}$ & $-0.0063^{+0.0057}_{-0.0059}$ \\

$K_1$ ($\rm{km \: s^{-1}}$) & $\mathcal{U}$(-$\infty$, $\infty$)  &  $19.61^{+0.26}_{-0.28}$ & $19.72^{+0.38}_{-0.62}$ & $19.75^{+0.10}_{-0.10}$ \\

$\gamma$ ($\rm{km \: s^{-1}}$) & $\mathcal{U}$(-$\infty$, $\infty$)  &  $-27.38^{+0.18}_{-0.20}$ & $-27.48^{+0.26}_{-0.50}$ & $-27.39^{+0.10}_{-0.11}$ \\

$\gamma_o$ ($\rm{km \: s^{-1}}$) & $\mathcal{U}$(-$\infty$, $\infty$)  &  ... & ... & $-0.03^{+0.17}_{-0.17}$ \\

$\sigma_{\rm{j, HIRES}}$ ($\rm{km \: s^{-1}}$) & $\mathcal{U}$(0, 1)  & ... & $0.308^{+0.546}_{-0.274}$  & $0.187^{+0.298}_{-0.161}$ \\

$\sigma_{\rm{j, TRES}}$ ($\rm{km \: s^{-1}}$) & $\mathcal{U}$(0, 1) & $0.49^{+0.60}_{-0.39}$  & ... & $0.327^{+0.397}_{-0.241}$ \\

$e$ & ... & $0.015^{+0.020}_{-0.011}$  & $0.029^{+0.071}_{-0.022}$ & $0.0093^{+0.0061}_{-0.0055}$  \\

$\omega$ (deg) & ... & $-33^{+76}_{-37}$ & $-52^{+103}_{-30}$ & $-50^{+41}_{-26}$  \\

\hline
\end{tabular}
\end{table}

\begin{figure}[!htb]
\centering
\includegraphics[width=.79\textwidth]{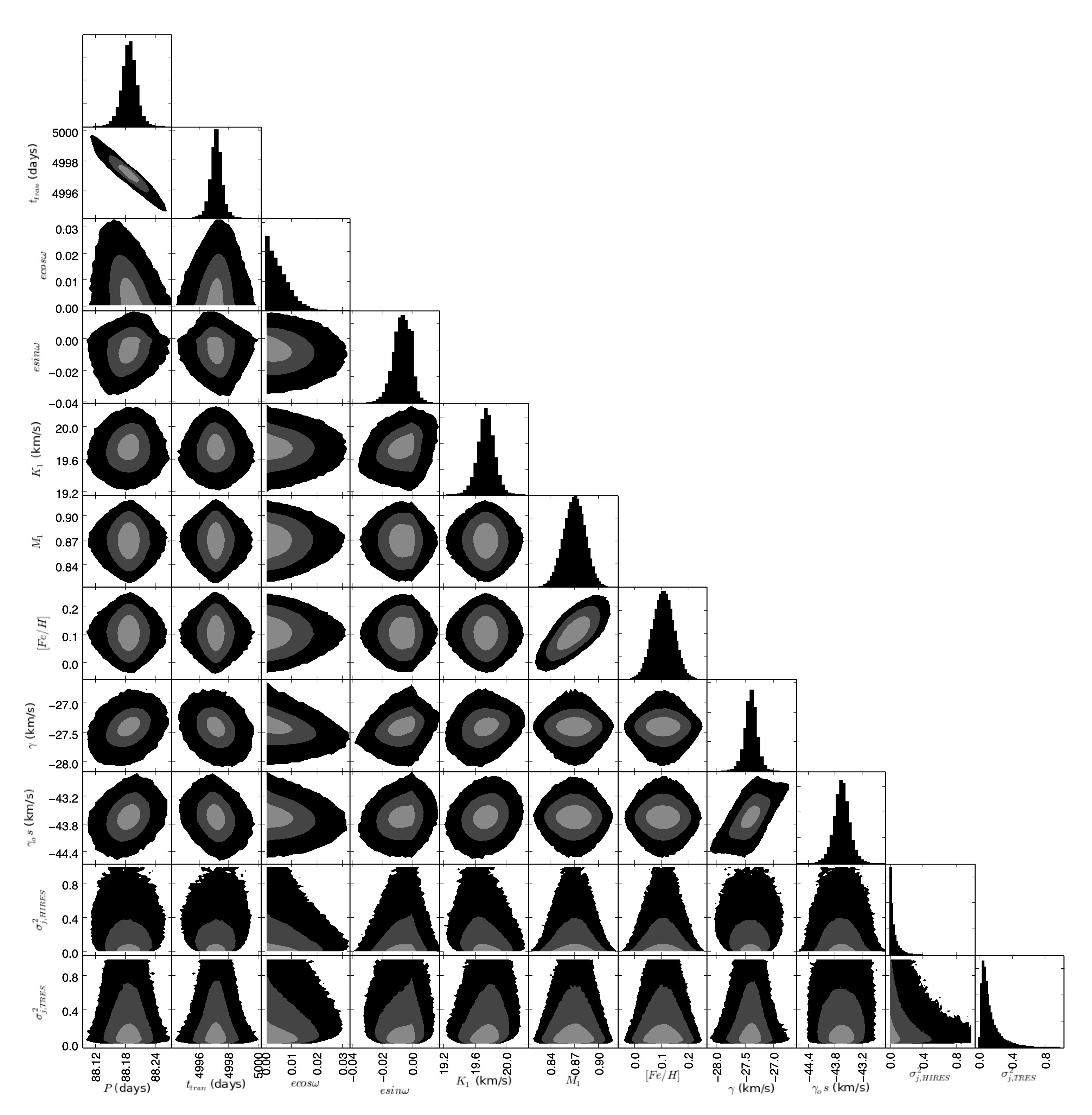}
\caption{Contour plots showing the $1 \sigma$, $2 \sigma$, and $3 \sigma$ constraints on pairs of parameters for the dynamical MCMC model constrained by stellar estimates of the G star from Brewer's analysis on HIRES data and radial velocities from both HIRES and TRES. Masses are all in units of solar masses.}
\label{fig: RVmass_Triangle}
\end{figure}

We then fit the spectroscopic data to a MCMC model including both radial velocity models and isochrone models. In order to be consistent with the microlensing models, we fit the stellar parameters for the primary (surface gravity, metallicity, and effective temperature) estimated by SPC, Brewer, and SpecMatch to the Padova PARSEC Isochrones, as was used in \cite{KruseAgol}. Using the Padova PARSEC isochrones \citep{PARSEC}, we can constrain predictions for the radius and the mass of the G star star. Padova PARSEC is a publicly available grid of stellar models that provides information on stars with parameters in the ranges: ages from $0.004 < t_1 < 12.59$ Gyr (spaced by 0.05 dex), metallicities from $-1.8 < [\rm{Fe/H}]_1 < 0.7$ (spaced by 0.1 dex), and masses from $0.1 < M_1 < 11.75 \, M_\odot$. Spacings in the isochrone model depend on age and metallicity and are adaptively chosen by the isochrone model. 

As done in the \cite{KruseAgol} models, we included the metallicity and the mass of the primary as free parameters in the model, for a total for 11 modelling parameters. Using the mass of the primary and the metallicity at each step, we determine the lifetime of the primary. We then interpolate all desired observables from the isochrone at each step, based on the input mass, age, and metallicity of the primary. We find the four bounding combinations of metallicity and age in the grid of the isochrone at the input mass, using a mass interpolation function. Then, we preform a bilinear interpolation of the four locations on the isochrone grid in order to determine the predicted value for the observables. Notably, using this model, we obtain predictions for the radius, surface gravity, and effective temperature of the G star. At each step, if a set of inputs into the Padova PARSEC isochrones falls outside the grid, we return a likelihood of negative infinity, as is done for nonphysical parameters throughout the MCMC modelling. The resulting corner plot for the MCMC model fit to both orbital and stellar evolutionary models using both TRES and HIRES velocities and the Brewer's spectroscopic estimates can be seen in Figure \ref{fig: RVmass_Triangle}.

We ran three MCMC models with 100,000 steps and 100 walkers for this Newtonian global model, with spectroscopy from both HIRES and TRES and with radial velocity and isochrone models. We ran an independent MCMC model for each of the SPC, Brewer, and SpecMatch estimates of G star parameters from HIRES spectroscopy. We threw out the first 2,000 steps as burn in for each MCMC model. We tested for convergence by enforcing that the number of independent draws was greater than 1,000 and determining the Gelman-Rubin statistic for each modeled parameter \citep{Ford, RadVel}. The maximum Gelman-Rubin value of the chains is 1.003. The median modeled MCMC parameters have a reduced chi-square of $\frac{\chi^2}{\rm{DOF}}_{SPC} = 0.94$, $\frac{\chi^2}{\rm{DOF}}_{Brewer} = 0.99$, and  $\frac{\chi^2}{\rm{DOF}}\footnote{The $\frac{\chi^2}{\rm{DOF}}$ for the SpecMatch MCMC model is greater than unity primarily due to the $\chi^2$ penalty from the difference between the median MCMC modeled surface gravity (4.431) and the SpecMatch prediction (4.62 $\pm$ 0.07).}_{SpecMatch} = 1.60$, respectively.

\begin{table} 
\centering
\caption{Dynamical MCMC Model Predictions} 
\label{tab: compare_RVmass_results}
\begin{tabular}{ c  c  c  c  c }
\hline
Parameter & Prior\footnote{\label{fn: priorLC} Priors adopted in the MCMC model. If no prior is listed, then the parameter is a derived parameter. $\mathcal{U}(x,y)$ denotes a uniform distribution between x and y. $\mathcal{N}(\mu, \sigma^2$) denotes a Gaussian distribution centered at $\mu$ with width of $\sigma$.} & SPC & Brewer & SpecMatch \\
\hline

$ M_{\rm{2}}$ ($M\textsubscript{\(\odot\)}$) & ... & $0.5220^{+0.0081}_{-0.0081}$ & $0.5122^{+0.0057}_{-0.0058}$ &$0.5207^{+0.0063}_{-0.0063}$ \\

$M_1$ ($M\textsubscript{\(\odot\)}$) &  $\mathcal{U}$(0, $\infty$) & $0.900^{+0.022}_{-0.022}$ & $0.870^{+0.013}_{-0.014}$ & $0.896^{+0.016}_{-0.015}$ \\

[Fe/H] &  $\mathcal{N}(\mu_{spec}, \sigma^2_{spec})$\footnote{\label{fn: specPrior} Gaussian prior around the spectroscopic estimates of the stellar parameters (effective temperature, metallicity, and surface gravity, respectively) from SPC, Brewer, and SpecMatch. Metallicity is a free parameter in the model. Effective temperature and surface gravity are derived parameters in the model. See Table \ref{tab: specPredictions} for spectroscopic estimates ($\mu_{\rm{spec}}$) and errors ($\sigma_{\rm{spec}}$).} & $0.178^{+0.078}_{-0.078}$ &  $0.109^{+0.039}_{-0.040}$ &  $0.144^{+0.039}_{-0.039}$\\

$T_{\rm{eff,1}}$ (K) & $\mathcal{N}(\mu_{spec}, \sigma^2_{spec})^{\rm{\ref{fn: specPrior}}}$ &  $5421^{+48}_{-48}$ & $5364^{+43}_{-43}$ & $5438^{+53}_{-55}$\\

$\log g$ & $\mathcal{N}(\mu_{spec}, \sigma^2_{spec})^{\rm{\ref{fn: specPrior}}}$ & $4.429^{+0.024}_{-0.029}$  &$4.464^{+0.015}_{-0.015}$ & $4.431^{+0.020}_{-0.024}$\\

$R_1$ ($R\textsubscript{\(\odot\)}$) & ... & $0.955^{+0.045}_{-0.037}$ & $0.902^{+0.023}_{-0.022}$ & $0.950^{+0.035}_{-0.029}$\\

\hline

\end{tabular}
\end{table}

We can also derive a modeled prediction for the white dwarf mass, $M_2$, independent of the photometric observations and microlensing models. From the initial fit to the occultations of KOI-3278, targeted as a planet candidate, the inclination was estimated as $89.6 ^\circ$ (see Table \ref{tab: compare_lc_spec_results}). Assuming inclination equals $90^\circ$ for this approximation, we can solve for $M_{\rm{2}}$ in Equation \ref{eq: WDeq}. The predicted stellar parameters for the MCMC model with dynamical and stellar evolutionary constraints, as well as the model parameters and priors, can be seen in Table \ref{tab: compare_RVmass_results}. From this dynamical and isochrone MCMC model constrained solely by spectroscopic observations and Brewer's stellar estimates, we predict a white dwarf mass of $M_{\rm{2}} = 0.5122^{+0.0057}_{-0.0058} \, M\textsubscript{\(\odot\)}$. For a detailed discussion regarding the  mass estimate of the white dwarf, see Section \ref{sec: discussion}.

\begin{equation}\label{eq: WDeq}
	K = \Big[ \frac{2 \pi G}{P (M_1 + M_2)^2} \Big] ^\frac{1}{3} \; \frac{M_2 \sin i}{\sqrt{1 - e^2}}
\end{equation}

\vspace{5mm}
\section{Joint Einsteinian and Newtonian Model} \label{sec: Joint}

We then created a joint Einsteinian microlensing and Newtonian radial velocity model to fit the photometric observations, the spectroscopic estimates of stellar parameters, and the spectroscopic radial velocities. 

In the joint model, we are able to remove all assumptions on the mass-radius relationship of the white dwarf. Doing so provides a test on mass-radius models for white dwarfs, as we independently model the white dwarf mass and radius. In order to do so, at each step of the MCMC model we solve for the white dwarf mass using Newtonian dynamical equations, as described in Section \ref{subsec: RV Model} and Equation \ref{eq: WDeq}. Next, we solve for the microlensing pulse height, ``h'', as a function of the primary radius, white dwarf radius, and Einstein radius of the white dwarf, using Equation \ref{eq: pulseHeight}. We can solve for the Einstein radius of the white dwarf throughout the orbital cycle by using Equation \ref{eq: Rein} \citep{Han}. The pulse height, or lensing magnification minus one ($A - 1$), is the difference between the microlensing magnification and the white dwarf occultation, and is in turn used in the Mandel-Agol procedure to fit the light curve \citep{MandelAgol},

\begin{equation} \label{eq: pulseHeight}
    h = A - 1 = \frac{2 R_E^2 - R_2^2}{R_1^2},
\end{equation}
where
\begin{equation} \label{eq: Rein}
    R_E = \sqrt{\frac{4 \, G \, M_2 \, a}{c^2}}.
\end{equation}

The joint model also allows us to remove all white dwarf evolution models and all assumptions on the initial to final mass relationship for white dwarfs. Without using these models to estimate the flux ratio of the two stars, we add the flux ratio of the white dwarf to the G star ($F_2/F_1$) as a free parameter in the joint model, as we can no longer constrain the flux of the white dwarf from white dwarf models.

In the joint microlensing and dynamical model, we have 15 fitted parameters: period ($P$), transit time ($t_{\rm{tran}}$), eccentricity ($e$) and longitude of periapsis of the G star ($\omega$) as $e \cos \omega$ and $e \sin \omega$, impact parameter ($b$), white dwarf radius ($R_2$), G star mass ($M_1$), metallicity of G star ($[\rm{Fe/H}]_1$), log age of the system, radial velocity semi-amplitude ($K$), center of mass velocity ($\gamma$), zero point offset between the center of mass velocities of the two spectra ($\gamma_o$), stellar jitter terms for the two sets of spectra squared ($\sigma^2_{\rm{j, HIRES}}$ and $\sigma^2_{\rm{j, TRES}}$), and the flux ratio of the Kepler photometry between the two stars ($F_2/F_1$). The modelling parameters and priors are listed in Table \ref{tab: Joint_model_results}.

The joint model should better constrain the impact parameter. This is because for purely Einsteinian photometric models, the duration of the pulse is a function of both the impact parameter and the velocity at the times of inferior and superior conjunction. These depend in an opposite manner on $e \sin{\omega}$. For an impact parameter of $\frac{1}{\sqrt{2}}$ the dependence on $e \sin{\omega}$ disappears at linear order. As our modeled impact parameter is close to this value, including radial velocity to help constrain $e \sin{\omega}$ should in turn improve our constrain on the impact parameter \citep{Carter, Winn}.

We ran three MCMC models with 100,000 steps and 50 walkers, and we threw out the first 20,000 steps as burn in. We ran an independent MCMC model for each of the SPC, Brewer, and SpecMatch estimates of G star parameters from the HIRES spectroscopy. We tested for convergence by enforcing that the number of independent draws was greater than 1,000 and determining the Gelman-Rubin statistic for each modeled parameter \citep{Ford, RadVel}. The maximum Gelman-Rubin value of the chains is 1.008. The median modeled MCMC parameters have a reduced chi-square of $\frac{\chi^2}{\rm{DOF}}_{SPC} = 1.03$, $\frac{\chi^2}{\rm{DOF}}_{Brewer} = 1.03$, and $\frac{\chi^2}{\rm{DOF}}_{SpecMatch} = 1.03$, respectively.

\begin{figure}[!htb]
\centering
\includegraphics[width=.97\textwidth]{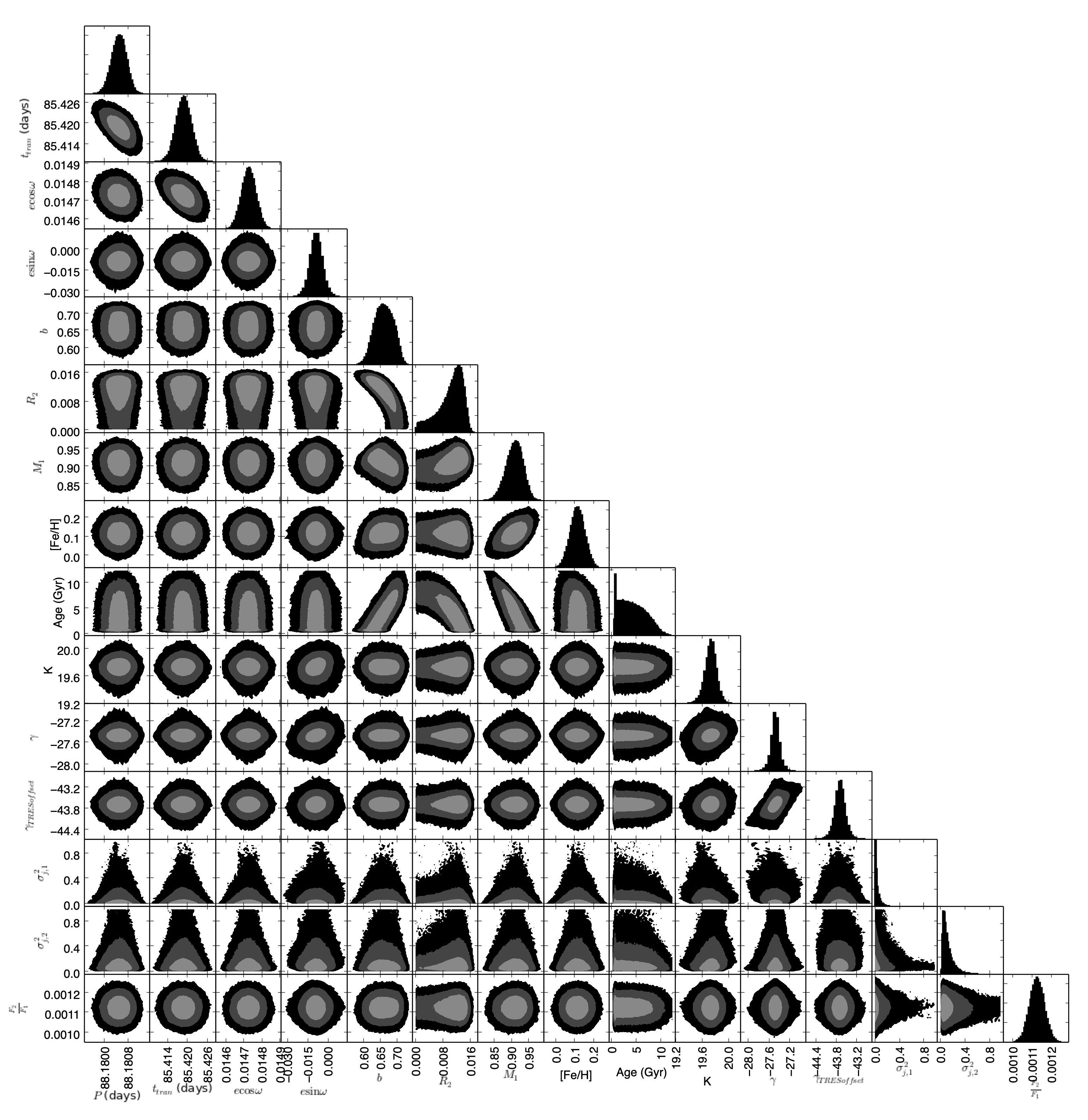}
\caption{Contour plots showing the $1 \sigma$, $2 \sigma$, and $3 \sigma$ constraints on pairs of parameters for the joint Einsteinian microlensing and Newtonian radial velocity MCMC model constrained by Kepler photometry, stellar estimates of the G star from Brewer's analysis on HIRES data, and radial velocities from both HIRES and TRES. Masses are all in units of solar masses.}
\vspace{2mm}
\label{fig: Joint_triangle}
\end{figure}

\begin{figure}[!htb]
\centering

\includegraphics[width=.72\textwidth]{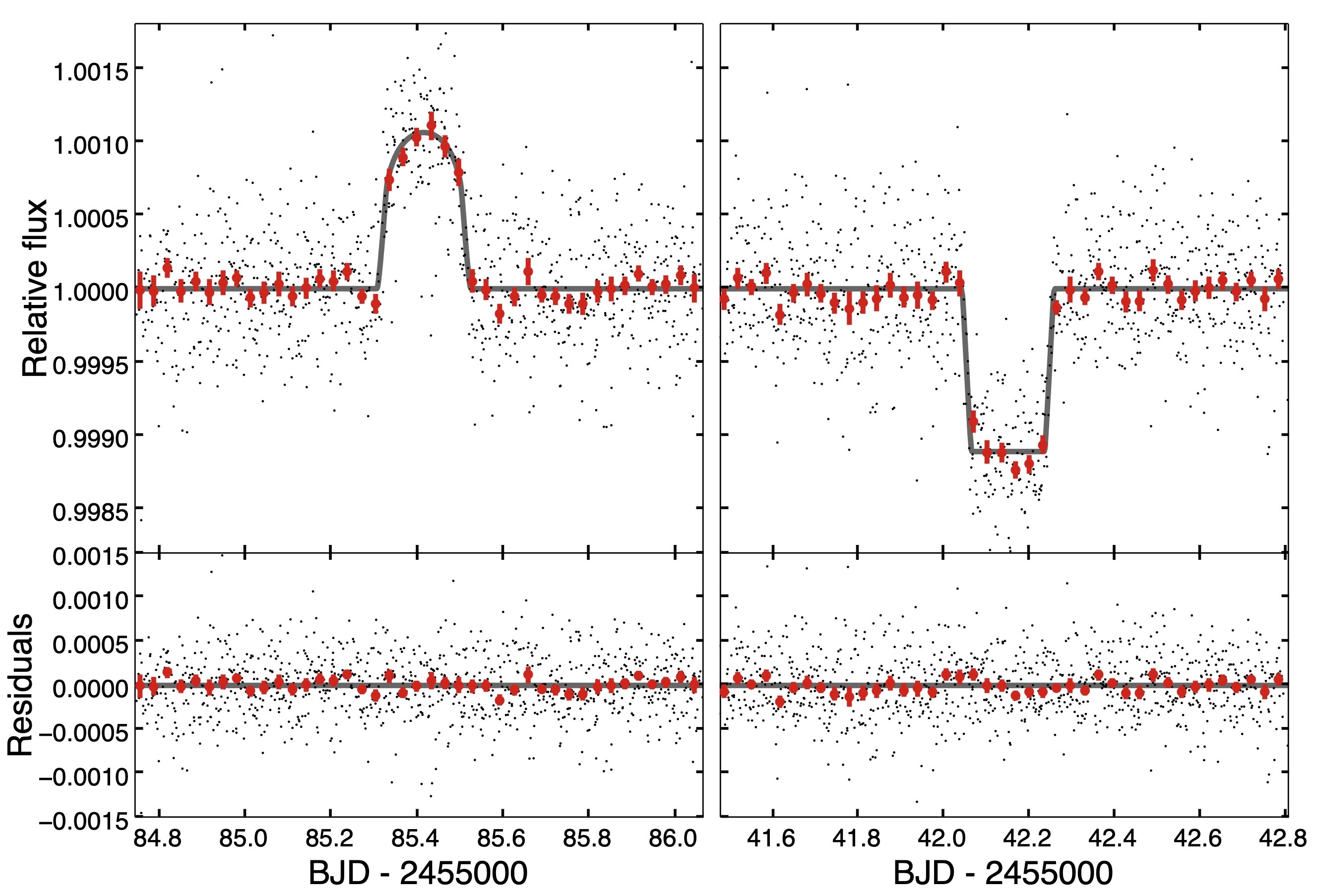}
\caption{Phase-folded light curves for KOI-3278.  The detrended Kepler fluxes are plotted as black points, while the red points and error bars show the fluxes in 45-minute bins.  The maximum-likelihood fit from the joint model using Brewer's stellar estimates from HIRES spectra is plotted as a continuous gray line for the microlensing pulses (left panels) and white-dwarf occultations (right panel).  The residuals from the model fit are plotted in the lower panels.}
\label{fig: Joint_Detrended_LC}

\bigskip

\includegraphics[width=.72\textwidth]{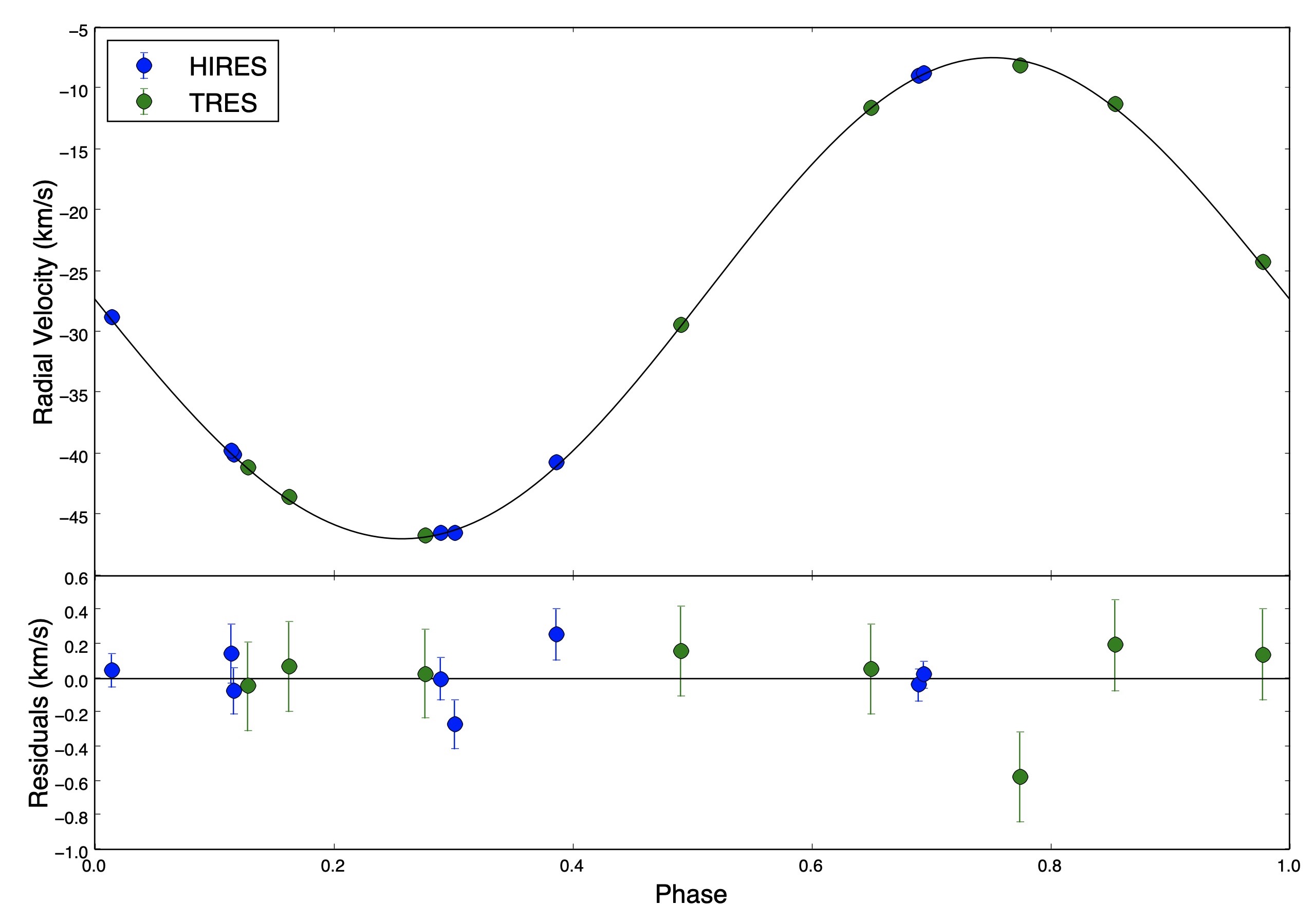}
\caption{TRES and HIRES radial velocity maximum-likelihood MCMC fit from the joint model using Brewer's stellar estimates from HIRES spectra. Blue data points are HIRES radial velocities and green data points are TRES radial velocities. RMS residual velocity of 0.19 $\rm{km \: s^{-1}}$. The uncertainties in radial velocity observations are the reported errors from the spectroscopic radial velocities added in quadrature with the MCMC modelled radial velocity jitter terms.}
\label{fig: JointRV}

\end{figure}

The corner plot for the MCMC model using Brewer's stellar estimates can be seen in Figure \ref{fig: Joint_triangle}, and the median modeled parameters are reported in Table \ref{tab: Joint_model_results}. The detrended and phase-folded Kepler photometry together with the maximum-likelihood joint model fit to the light curve can be seen in Figure \ref{fig: Joint_Detrended_LC}. The maximum-likelihood joint model fit to the radial velocity observations can be seen in Figure \ref{fig: JointRV}.  Using Brewer's stellar estimates as priors on MCMC model, our joint microlensing and radial velocity prediction for the white dwarf mass is $0.5250^{+0.0082}_{-0.0089} \, M\textsubscript{\(\odot\)}$. For a longer discussion regarding the  mass estimate of the white dwarf, see Section \ref{sec: discussion}.

\begin{table} 
\centering
\caption{Parameters from the Joint Microlensing and Radial Velocity Model}
\label{tab: Joint_model_results}
\begin{tabular}{ c  c  c  c  c }
\bigskip\\

\multicolumn{5}{c}{G Star}\\
\hline
Parameter & Prior\footnote{\label{fn: priorLC} Priors adopted in the MCMC model. If no prior is listed, then the parameter is a derived parameter. $\mathcal{U}(x,y)$ denotes a uniform distribution between x and y. $\mathcal{N}(\mu, \sigma^2$) denotes a Gaussian distribution centered at $\mu$ with width of $\sigma$.} & SPC & Brewer & SpecMatch \\
\hline 

$M_1$ ($M\textsubscript{\(\odot\)}$) & $\mathcal{U}$(0, $\infty$) & $0.951^{+0.030}_{-0.032}$ & $0.911^{+0.023}_{-0.026}$  & $0.955^{+0.024}_{-0.026}$ \\

$R_1$ ($R\textsubscript{\(\odot\)}$) & ... & $0.896^{+0.027}_{-0.029}$ &  $0.861^{+0.028}_{-0.023}$ & $0.890^{+0.026}_{-0.025}$  \\

[Fe/H]$_1$ & $\mathcal{N}(\mu_{spec}, \sigma^2_{spec})$\footnote{\label{fn: specPrior} Gaussian prior around the spectroscopic estimates of the stellar parameters (effective temperature, metallicity, and surface gravity, respectively) from SPC, Brewer, and SpecMatch. Metallicity is a free parameter in the model. Effective temperature and surface gravity are derived parameters in the model. See Table \ref{tab: specPredictions} for spectroscopic estimates ($\mu_{\rm{spec}}$) and errors ($\sigma_{\rm{spec}}$).} & $0.208^{+0.079}_{-0.079}$ &  $0.118^{+0.040}_{-0.040}$  & $0.155^{+0.040}_{-0.040}$ \\

Age$_1$ (Gyr) & $\mathcal{N}(0.89, 0.15^2)$\footnote{If age of the star is less than the spin-down age of star, Gaussian prior around spin-down age = $0.89 \pm 0.15$ Gyr \citep{KruseAgol}.} & $3.5^{+2.6}_{-2.1}$ &  $4.3^{+3.2}_{-2.6}$ & $2.7^{+2.4}_{-1.6}$ \\

$T_{\rm{eff,1}}$ (K) & $\mathcal{N}(\mu_{spec}, \sigma^2_{spec})^{\rm{\ref{fn: specPrior}}}$ & $5436^{+50}_{-50}$ & $5384^{+45}_{-44}$ & $5484^{+58}_{-58}$ \\

$\log g_1$  & $\mathcal{N}(\mu_{spec}, \sigma^2_{spec})^{\rm{\ref{fn: specPrior}}}$ &  $4.509^{+0.028}_{-0.028}$ & $4.525^{+0.028}_{-0.035}$ & $4.518^{+0.022}_{-0.029}$ \\
\hline

\bigskip\\
\multicolumn{5}{c}{White Dwarf}\\
\hline 
Parameter & Prior$^{\rm{\ref{fn: priorLC}}}$ & SPC & Brewer & SpecMatch \\
\hline 

$M_2$ ($M\textsubscript{\(\odot\)}$)  & ... &  $0.5379^{+0.0100}_{-0.0107}$  & $0.5250^{+0.0082}_{-0.0089}$ & $0.5392^{+0.0081}_{-0.0088}$\\

$R_2$ ($R\textsubscript{\(\odot\)}$) & $\mathcal{U}$(0, $\infty$) &  $0.0089^{+0.0034}_{-0.0051}$ & $0.0111^{+0.0026}_{-0.0048}$ & $0.0099^{+0.0027}_{-0.0049}$ \\

$R_E$ ($R\textsubscript{\(\odot\)}$) &  ... & $0.02094^{+0.00028}_{-0.00031}$ & $0.02056^{+0.00023}_{-0.00026}$  & $0.02097^{+0.00023}_{-0.00025}$ \\

\hline

\bigskip\\
\multicolumn{5}{c}{Binary System}\\
\hline
Parameter & Prior$^{\rm{\ref{fn: priorLC}}}$  & SPC & Brewer & SpecMatch \\
\hline 

$P$ (days)  & $\mathcal{U}$(0, $\infty$) & $88.18053^{+0.00025}_{-0.00026}$  & $88.18052^{+0.00025}_{-0.00026}$& $88.18052^{+0.00025}_{-0.00026}$  \\

$t_{\rm{tran}}$ (BJD $- \, 2,455,000$) & $\mathcal{U}$(-$\infty$, $\infty$) & $85.4190^{+0.0023}_{-0.0023}$ & $85.4190^{+0.0023}_{-0.0023}$  & $85.4190^{+0.0023}_{-0.0023}$  \\

$e\: \rm{cos} \: \omega$ & $\mathcal{U}$(-1, 1) &  $0.014730^{+0.000041}_{-0.000042}$ & $0.014730^{+0.000041}_{-0.000041}$ & $0.014729^{+0.000041}_{-0.000041}$ \\

$e\: \rm{sin} \: \omega$  & $\mathcal{U}$(-1, 1) & $-0.0082^{+0.0047}_{-0.0048}$ &   $-0.0083^{+0.0048}_{-0.0048}$ & $-0.0081^{+0.0048}_{-0.0048}$ \\

$b$   & $\mathcal{U}$(-$\infty$, $\infty$) & $0.686^{+0.023}_{-0.029}$  & $0.663^{+0.030}_{-0.029}$  & $0.680^{+0.024}_{-0.025}$\\

$e$ &  ... & $0.0169^{+0.0028}_{-0.0017}$ & $0.0169^{+0.0028}_{-0.0017}$ &  $0.0168^{+0.0027}_{-0.0017}$ \\

$\omega$ (deg)  & ... & $-29^{+16}_{-12}$ &  $-29^{+16}_{-12}$ &  $-29^{+16}_{-12}$ \\

$a$ (AU)  & ... & $0.4426^{+0.0039}_{-0.0043}$ &  $0.4373^{+0.0031}_{-0.0035}$  & $0.4431^{+0.0031}_{-0.0034}$ \\

$a/R_1$ &  ... & $106.2^{+3.5}_{-3.1}$   &  $109.3^{+3.2}_{-3.9}$  & $107.3^{+2.9}_{-3.2}$  \\

$i$ (deg) & ... & $89.630^{+0.026}_{-0.023}$  &  $89.653^{+0.024}_{-0.029}$ & $89.637^{+0.022}_{-0.024}$ \\

$K_1$ ($\rm{km \: s^{-1}}$) & $\mathcal{U}$(-$\infty$, $\infty$)  & $19.744^{+0.085}_{-0.089}$ & $19.742^{+0.084}_{-0.090}$  & $19.743^{+0.086}_{-0.091}$  \\

$\gamma$ ($\rm{km \: s^{-1}}$) & $\mathcal{U}$(-$\infty$, $\infty$) & $-27.461^{+0.080}_{-0.081}$ & $-27.463^{+0.079}_{-0.081}$ & $-27.462^{+0.081}_{-0.083}$ \\

$\gamma_o$ ($\rm{km \: s^{-1}}$) & $\mathcal{U}$(-$\infty$, $\infty$) & $-0.07^{+0.15}_{-0.15}$ & $-0.07^{+0.15}_{-0.15}$ & $-0.07^{+0.15}_{-0.15}$ \\

$\sigma_{\rm{j, HIRES}}$ ($\rm{km \: s^{-1}}$) & $\mathcal{U}$(0, 1) & $0.167^{+0.245}_{-0.145}$ & $0.167^{+0.247}_{-0.145}$  & $0.170^{+0.247}_{-0.145}$ \\

$\sigma_{\rm{j, TRES}}$ ($\rm{km \: s^{-1}}$) & $\mathcal{U}$(0, 1) & $0.324^{+0.359}_{-0.226}$  & $0.326^{+0.352}_{-0.228}$ & $0.322^{+0.349}_{-0.226}$ \\

$u_1$ & ... & $0.477^{+0.012}_{-0.012}$  & $0.4797^{+0.0094}_{-0.0095}$ & $0.463^{+0.012}_{-0.012}$ \\

$u_2$ & ... & $0.2254^{+0.0075}_{-0.0075}$  & $0.2365^{+0.0060}_{-0.0060}$ & $0.2218^{+0.0078}_{-0.0077}$ \\

Magnification - 1 & ... & $0.000989^{+0.000039}_{-0.000041}$  &   $0.000977^{+0.000042}_{-0.000043}$ & $0.000988^{+0.000040}_{-0.000042}$ \\

$F_2/F_1$ & ... & $0.001128^{+0.000039}_{-0.000038}$ & $0.001128^{+0.000039}_{-0.000039}$ & $0.001128^{+0.000039}_{-0.000039}$\\

\hline
\end{tabular}
\end{table}

\clearpage
\section{Discussion} \label{sec: discussion}
\subsection{White Dwarf Mass}

\begin{figure}[!htb]
\centering
\includegraphics[width=.77\textwidth]{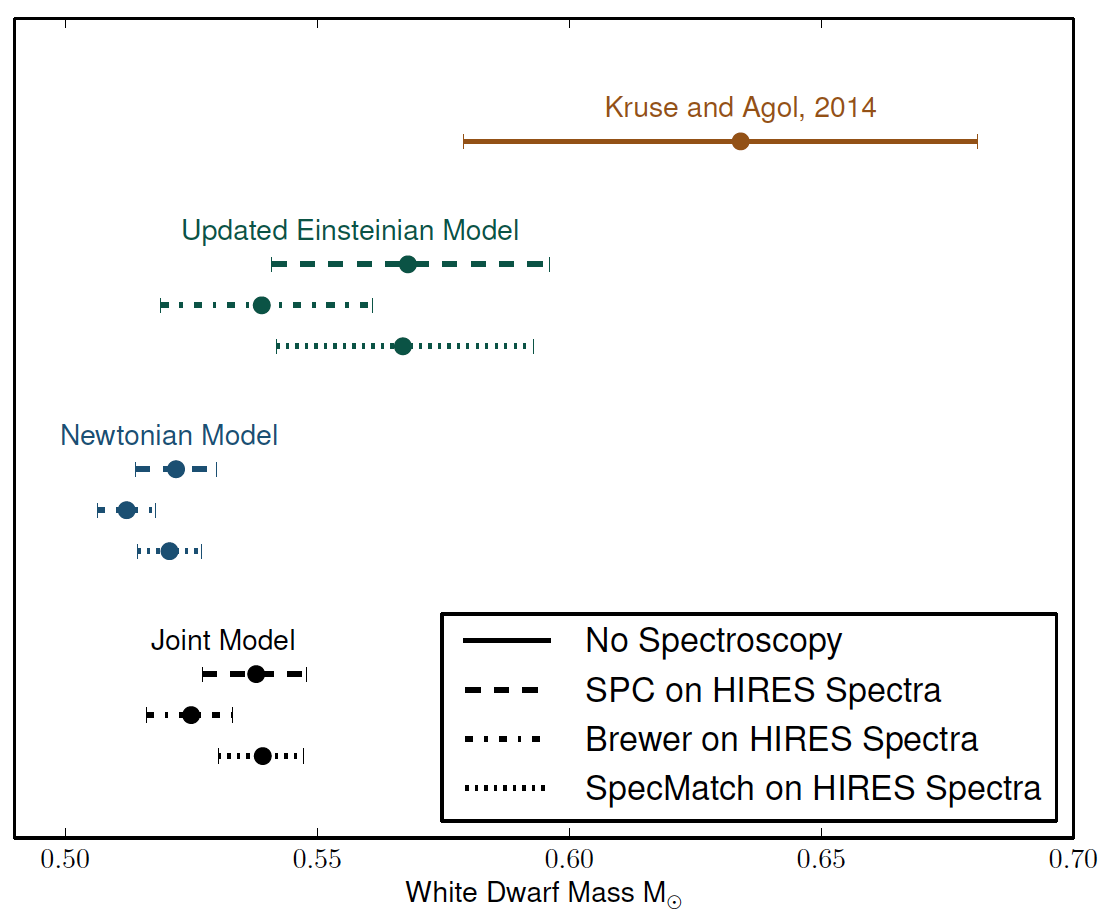}
\caption{White dwarf mass predictions from purely photometric \citet{KruseAgol} Einsteinian microlensing model, the updated photometric and spectroscopic Einsteinian microlensing models, the purely spectroscopic Newtonian radial velocity models, and the photometric and spectroscopic joint Einsteinian microlensing and Newtonian radial velocity models.}
\vspace{2mm}
\label{fig: WDmass}
\end{figure}

We believe that the biggest limiting factor in our ability to precisely model the mass of the the white dwarf stems from our ability to constrain the isochrone models. This difficulty arises from two main factors: (1) determining stellar parameter predictions from spectroscopy and (2) applying isochrone models to a stellar binary in order to predict the mass and radius of the primary. The stellar parameters estimates from SPC, Brewer, and SpecMatch analyses on HIRES differ at $2.0\%$ level for the effective temperature, at the $17.5 \%$ level for surface gravity, and at the $25.9 \%$ level for the metallicity. These differences affect our ability to accurately constrain the mass and radius of the primary. In turn, this diminishes our ability to precisely predict the mass of the white dwarf, and plays a role in the differences between the white dwarf mass predictions. The differences in white dwarf mass predictions, based solely on the spectroscopic estimates of stellar parameters using SPC, Brewer, and SpecMatch analysis of HIRES spectra, can be seen in Figure \ref{fig: WDmass}. In addition, we constrain the mass and radius of the G star using the Padova PARSEC isochrone models; however, there is evidence that stellar binaries often fall on unusual locations in stellar evolutionary models \citep{Kawahara}. Therefore, constraining stellar parameters using these isochrones also reduces our ability to precisely and accurately model the stellar parameters.

Our updated microlensing model (see Section \ref{sec: LC}) uses photometric constraints on the light curve model and spectroscopic constraints on the isochrone model. These models therefore incorporate some Newtonian physics in the spectroscopic predictions of $T_{\rm{eff}}$, $\log g$, and [Fe/H], however they are independent of the orbital modelling predictions - and the white dwarf predicted mass is predominantly Einsteinian. Our three independent MCMC models, using the different sets of stellar parameter estimates as Gaussian priors, resulted in the following three white dwarf mass predictions: the model using SPC analysis predicted a white dwarf mass of $0.568^{+0.028}_{-0.027}$ $M\textsubscript{\(\odot\)}$, the  model using Brewer's analysis predicted a white dwarf mass of $0.539^{+0.022}_{-0.020}$ $M\textsubscript{\(\odot\)}$, while the model using SpecMatch analysis predicted a white dwarf mass of $0.567^{+0.026}_{-0.025}$ $M\textsubscript{\(\odot\)}$. These three models, which only vary in the priors set on the primary star effective temperature, surface gravity, and metallicity based on the SPC, Brewer, and SpecMatch analyses, differ in the mass predictions as much as $5.4 \%$. Using an Einsteinian microlensing model without any spectroscopy, \cite{KruseAgol} predicted a white dwarf mass of $0.634^{+0.047}_{-0.055}$ $M\textsubscript{\(\odot\)}$. The difference between these two models is predominantly due to the difference in the mass and radius prediction of the primary. This stems from the updated estimates of metallicity, surface gravity, and effective temperature of the primary, determined from spectroscopy.

The purely dynamical models (see Section \ref{sec: RV}) presented in this paper rely solely on spectroscopic constraints on the Padova PARSEC models and an orbital solution to the radial velocities. It is true that at high enough velocities, there is a special relativistic correction to the Doppler shift of the spectroscopy. However, at velocities on the order of tens of $\rm{km \: s^{-1}}$ this relativistic correction is negligible. Therefore, the dynamical model follows from purely Newtonian predictions. The Newtonian dynamical MCMC models, using the different sets of stellar parameter estimates as Gaussian priors, resulted in the following three white dwarf mass predictions: the model using SPC analysis predicted a white dwarf mass of $0.5220^{+0.0081}_{-0.0081}$ $M\textsubscript{\(\odot\)}$, the model using Brewer's analysis predicted a white dwarf mass of $0.5122^{+0.0057}_{-0.0058}$ $M\textsubscript{\(\odot\)}$, while the model using SpecMatch analysis predicted a white dwarf mass of $0.5207^{+0.0063}_{-0.0063}$ $M\textsubscript{\(\odot\)}$. These three models, which only vary in the priors set on the primary star effective temperature, surface gravity, and metallicity based on the SPC, Brewer, and SpecMatch analyses, differ as much as $1.9 \%$.

The independent Einsteinian microlensing model and the Newtonian dynamical model predict a white dwarf mass companion that differ by $8.8 \%$ using the SPC stellar estimates, $5.2 \%$ using Brewer's stellar estimates, and $8.9 \%$ using the SpecMatch stellar estimates. 

The joint Einsteinian microlensing and Newtonian dynamical model (see Section \ref{sec: Joint}) used photometric observations, spectroscopic radial velocities, and the three sets of spectroscopic estimates of stellar parameters in order to model the stellar binary. Our three independent MCMC models, using the different sets of stellar parameter estimates as Gaussian priors, resulted in the following three white dwarf mass predictions: the model using SPC analysis predicted a white dwarf mass of $0.5379^{+0.0100}_{-0.0107}$ $M\textsubscript{\(\odot\)}$, the  model using Brewer's analysis predicted a white dwarf mass of $0.5250^{+0.0082}_{-0.0089}$ $M\textsubscript{\(\odot\)}$, while the model using SpecMatch analysis predicted a white dwarf mass of $0.5392^{+0.0081}_{-0.0088}$ $M\textsubscript{\(\odot\)}$. These three models, which only vary in the priors set on the primary star effective temperature, surface gravity, and metallicity based on the SPC, Brewer, and SpecMatch analyses, differ as much as $2.7 \%$.

The white dwarf mass predictions and uncertainties from the original \citet{KruseAgol} Einsteinian model, the updated Einsteinian model with spectroscopic constraints on the isochrones, the Newtonian model, and the joint model can be seen in Figure \ref{fig: WDmass}. As Brewer's spectroscopic estimates of the primary star parameters had the smallest errors, we believe this is our best estimate of the mass of the white dwarf companion in the Einsteinian model, the Newtonian model, and the joint model. As such the Brewer median MCMC modeled parameters, and 1$\sigma$ uncertainties are consistently reported in all figures and text.

\subsection{Mass-Radius Relationship of White Dwarfs}

\begin{figure}[!htb]
\centering
\includegraphics[width=\textwidth]{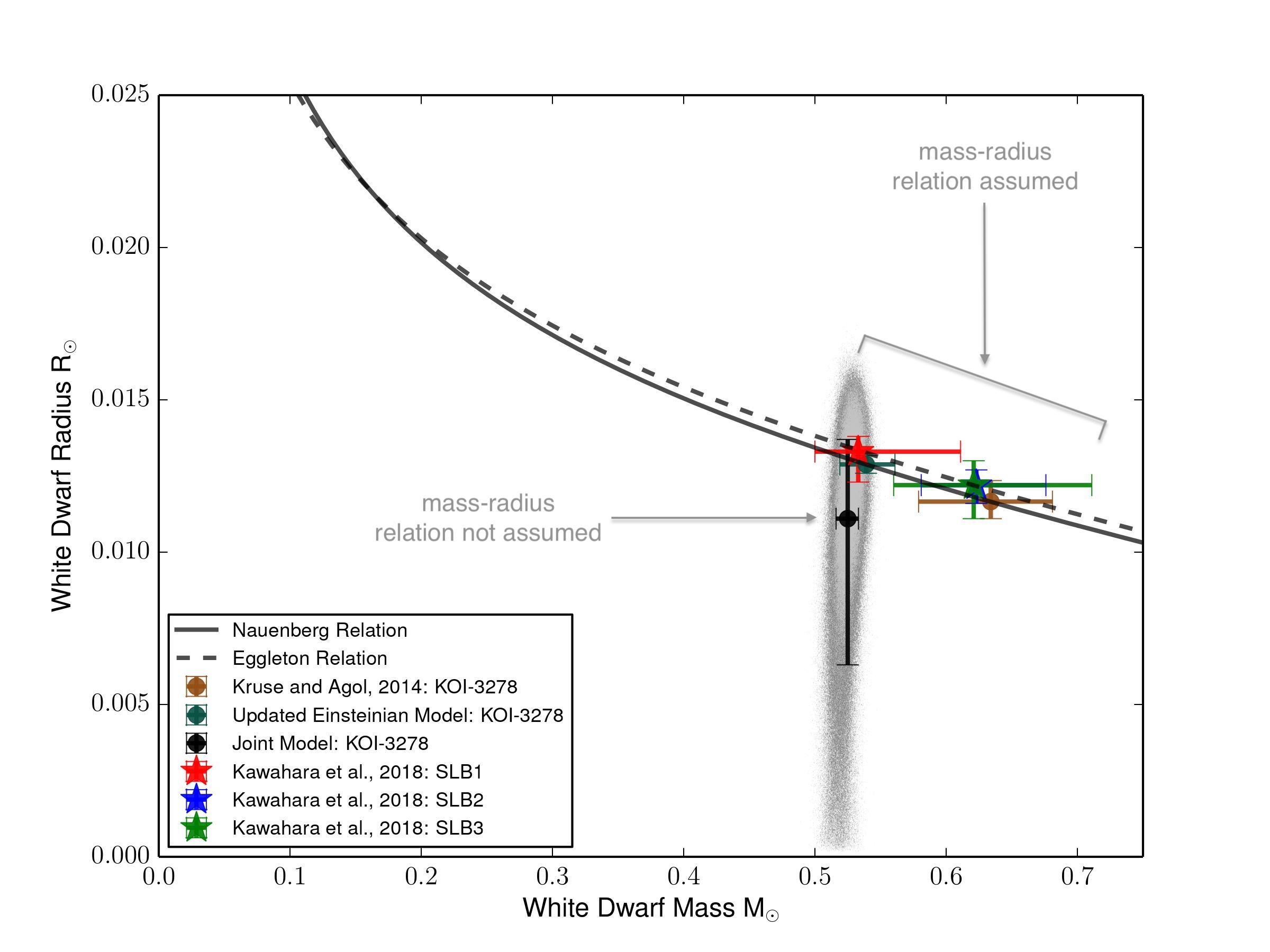}
\caption{White dwarf mass and radius predictions and uncertainties from the purely photometric \citet{KruseAgol} Einsteinian model, the updated photometric and spectroscopic Einsteinian model using Brewer's stellar estimates of HIRES spectroscopy, and the photometric and spectroscopic joint Einsteinian and Newtonian model using Brewer's stellar estimates of HIRES spectroscopy for KOI-3278 (circular data points). Contour plot shows the $1 \sigma$ and $2 \sigma$ constraints on the mass and radius from the joint MCMC model using Brewer's stellar estimates. Also includes the white dwarf mass and radius predictions and uncertainties constrained by a joint light curve and RV MCMC model for three self-lensing binaries as presented in \citet{Kawahara} (star shaped data points). The black solid line shows the Nauenberg mass-radius relation, as described by Equation \ref{eq: Nauenberg} and used to determine the radii in the \citet{KruseAgol} model and the updated Einsteinian model of KOI-3278. The black dashed line shows the Eggleton mass-radius relation, as described by Equation \ref{eq: Eggleton} and used to determine the radii in the \citet{Kawahara} models. The joint model of KOI-3278 uses no mass-radius relation while all other mass-radius measurements stem from an assumed mass-radius relation. The radius errors on all models except the joint model are from a propagation of the errors on the mass through an assumed mass-radius relation.}
\vspace{2mm}
\label{fig: WDmassVradius}
\end{figure}

As discussed previously (see Section \ref{sec: LC}), in the Einsteinian microlensing model we must adopt a mass-radius relationship for the white dwarf in the model. Similarly to \citet{KruseAgol}, in the updated Einsteinian microlensing model we used the Nauenberg relation for the zero-temperature white dwarf (see Equation \ref{eq: Nauenberg}). However, in the joint model, we are able to remove this assumption (see Section \ref{sec: Joint}). In so doing, we constrain the white dwarf mass and radius independent of any mass-radius relationship. Figure \ref{fig: WDmassVradius} shows the predicted mass and radius for the white dwarf in KOI-3278 from \citet{KruseAgol}, our updated Einsteinian model, and our joint model. Figure \ref{fig: WDmassVradius} also shows the three mass and radius predictions from the three self lensing binaries in \citet{Kawahara}. \citet{Kawahara} used the Eggleton mass-radius relation to derive the radius of the white dwarf. The Eggleton mass-radius relation can be seen in Equation \ref{eq: Eggleton}, where $M_{Ch} = 1.454 \, M\textsubscript{\(\odot\)}$ is the Chandrasekhar mass and $M_p = 0.00057 \, M\textsubscript{\(\odot\)}$ is a constant  \citep{VerbuntRappaport}. Figure \ref{fig: WDmassVradius} also includes the Nauenberg mass-radius relation for the zero temperature white dwarf and the Eggleton mass-radius relation for white dwarfs.

\begin{equation} \label{eq: Eggleton}
R_2 = 0.0114 \, \Bigg[ {\Big(\frac{M_2}{M_{Ch}}\Big)^{-\frac{2}{3}} - \Big(\frac{M_2}{M_{Ch}}\Big)^\frac{2}{3}} \Bigg] ^{\frac{1}{2}} \, \Bigg[1 \, + \, 3.5 \Big(\frac{M_2}{M_{p}}\Big)^{-\frac{2}{3}} + \Big(\frac{M_2}{M_{p}}\Big)^{-1} \Bigg] ^{-\frac{2}{3}}
\end{equation}

The joint model is the only point that can potentially constrain the relation itself as it does not rely on a mass-radius relation assumption. The other five data points on Figure \ref{fig: WDmassVradius} are not independent measurements of the mass and radius, as they assume a relation (either Nauenberg or Eggleton) and the error on the radius is a propagation of the error on the mass through the assumed relation. The results of the joint model suggest that the white dwarf relations function as an upper limit on the radius of the white dwarf.  This can be interpreted as the effect of constraining the mass and radius of the white dwarf using both Einsteinian lensing models and Newtonian dynamical models. Specifically, as the radius of the white dwarf increases, the mass of the white dwarf must also increase in order to maintain the same pulse height from the lensing equations. The mass of the white dwarf is constrained by the Newtonian model, and thus as the mass of the white dwarf increases it eventually comes in conflict with the radial velocity observations.

Follow-up studies of KOI-3278 could help to more precisely constrain the radius of the white dwarf. In our joint model, the radius of the white dwarf is poorly constrained because the pulse is dominated by the lensing effect, which is a mass dominant effect, and the white dwarf occultation contributes little to the pulse portion of the light curve. Observations of a spectrum of the secondary eclipse, in the ultraviolet (UV), would allow for a more precise constraint on the white dwarf radius and thus a test of the white dwarf mass-radius relations. UV observations of the secondary eclipse, in conjunction with white dwarf models, would also provide a more precise estimate of the effective temperature of the white dwarf.

\subsection{Parallax with Gaia DR2}
The parallax prediction ($\pi$) from \cite{KruseAgol}, $1.237^{+0.079}_{-0.053}$ milli-arc seconds, agree at the 1$\sigma$ level with the Gaia DR2 observations, $1.2697^{+0.0218}_{-0.0218}$ milli-arc seconds \citep{Gaia}. Had we kept distance as a free parameter in the system, we would have set a prior on the parallax with the Gaia DR2 observation. However, we removed distance as a free parameter, as we were able to constrain the isochrone models with the spectroscopic estimates of stellar primary parameters. Including distance would require assumptions on the dust distribution, which we decided to remove.

\subsection{What's Next for Self-Lensing Binaries?}

Gaia can be used in order to detect similar systems to KOI-3278 not in an edge-on configuration and hence not showing photometric variability due to eclipses and lensing. Through analyzing reflex motion of the G star around the white dwarf center of mass, $\alpha_1$, we can detect these stellar binaries. Assuming a $1\%$ geometric lensing probability of KOI-3278, we expect about 100 of these objects in Kepler target stars \citep{KruseAgol}. Data from TESS \citep{TESS} is likely to reveal binary systems where a black hole companion self-lenses its primary. Black holes or neutron stars in binaries are more likely to lens their primary with periods that will be observable by TESS in individual, 27.4 day, sectors \citep{MasudaHotokezaka}. White dwarf self-lensing binaries could potentially be found near the ecliptic poles in TESS observations, where sectors overlap to allow for observing signals with significantly larger periods.

To our knowledge, this is the first time that there has been an independent Newtonian radial velocity and Einsteinian microlensing prediction for a white dwarf mass. Previous binary systems have been modelled using joint microlensing and radial velocity models, but these systems have not had independent models for comparison of the predicted white dwarf mass \citep{Yee, Kawahara}. Future work modeling white dwarf masses and radii independently can provide a better understanding of the mass-radius relationship for white dwarfs.


\section{Conclusion}
Using estimates on primary star metallicity, surface gravity, and temperature from spectroscopic observations as constraints, we present an updated microlensing model of the self-lensing binary, KOI-3278. The updated Einsteinian microlensing model, using Brewer's stellar estimates, predicts a white dwarf mass of $0.539^{+0.022}_{-0.020} \, M\textsubscript{\(\odot\)}$. We then produce an independent dynamical model fit to radial velocities taken from a single-lined orbital solution to spectroscopic observations of KOI-3278. We find that the Newtonian dynamical model, using Brewer's stellar estimates, predicts a white dwarf mass of $0.5122^{+0.0057}_{-0.0058} \, M\textsubscript{\(\odot\)}$. These Einsteinian and Newtonian predictions for the white dwarf mass differ by $5.2 \%$. This agreement is encouraging but far from definitive. We then present a joint Einsteinian microlensing and Newtonian dynamical model of KOI-3278, which allows us to remove all white dwarf evolutionary models as well as white mass-radius assumptions from the MCMC model. The joint model, using Brewer's stellar estimates, predicts a white dwarf mass of $0.5250^{+0.0082}_{-0.0089} \, M\textsubscript{\(\odot\)}$. We compare the independent mass and radius predictions from the joint model against the Nauenberg and Eggleton mass-radius relations for white dwarfs. We discuss that these mass-radius relations appear to function as upper limits on the radius of the white dwarf. Finally, we discuss how future UV observations of the spectrum of the secondary eclipse could provide a tighter constraint on the radius of the white dwarf and thus a test for white dwarf mass-radius relations.

\acknowledgments
We thank George Zhou, Joseph Rodriguez, Allyson Bieryla, Jason Eastman, and Stephanie Douglas for stimulating conversations and guidance.

\software{
\texttt{Brewer} \citep{Brewer2016}, $\,$
\texttt{emcee} \citep{emcee}, $\,$
\texttt{matplotlib} \citep{matplotlib}, $\,$
\texttt{numpy} \citep{numpy}, $\,$
\texttt{scipy} \citep{scipy}, $\,$
\texttt{SPC} \citep{Buchhave2012}, $\,$
\texttt{SpecMatch} \citep{Petigura17b}
.}



\begin{singlespace}
\bibliography{KOI3278}
\bibliographystyle{plain}
\end{singlespace}





\appendix
\section{Maximum-Likelihood MCMC Modeled Parameters}
We report the maximum-likelihood values of the modeled parameters from all the MCMC models. Table \ref{tab: Best-compare_lc_spec_results} shows the maximum-likelihood parameters from the Einsteinian microlensing model. Table \ref{tab: Best-RVfits} shows the maximum-likelihood orbital parameters from the Newtonian radial velocity model. Table \ref{tab: Best-compare_RVmass_results} shows the additional maximum-likelihood parameters from the Newtonian dynamical model including isochrone fitting and the spectroscopic estimates of G star parameters. Table \ref{tab: Best-Joint_model_results} shows the maximum-likelihood parameters from the joint Einsteinian microlensing and Newtonian dynamical model. In all models, we modeled $e \cos\omega$ and $e \sin\omega$, however, we also report the derived maximum-likelihood values of e and $\omega$.

\setcounter{table}{0}
\renewcommand{\thetable}{A\arabic{table}}

\begin{table}[h]
\centering
\caption{Maximum-Likelihood Parameters from the Microlensing Model}
\label{tab: Best-compare_lc_spec_results}
\begin{tabular}{ c  c  c  c }

\hline
Parameter & SPC & Brewer & SpecMatch \\
\hline 

$P$ (days) & $88.18058$ & $88.18052$  & $88.18058$ \\

$t_{tran}$ (BJD $- \, 2,455,000$) & $85.4181$ & $85.4186$  & $85.4184$  \\

$e \cos\omega$  & $0.01473$ & $0.014733$ &  $0.01473$  \\

$e \sin\omega$  & $0.003$ & $-0.004$ &  $-0.005$  \\

$b$  & $0.678$ &  $0.65$ & $0.676$ \\

$M_{2,init}$ ($M\textsubscript{\(\odot\)}$) & $1.64$ & $1.51$ & $1.72$\\

$M_2$ ($M\textsubscript{\(\odot\)}$) & $0.573$ & $0.547$ & $0.566$ \\

$M_1$ ($M\textsubscript{\(\odot\)}$) & $0.952$ & $0.916$ & $0.954$ \\

[Fe/H]  & $0.192$ & $0.119$  & $0.171$  \\

Age (Gyr) & $3.31$ & $4.0$ & $3.0$ \\

$e$  & $0.015$ & $0.015$ & $0.015$ \\

$\omega$ (deg)  & $10.0$ & $-15.0$ & $-18.0$  \\

\hline

\end{tabular}
\end{table}

\begin{table}[h]
\centering
\caption{Maximum-Likelihood Orbital Parameters from the Newtonian Radial Velocity Model}
\label{tab: Best-RVfits}
\begin{tabular}{ c  c  c  c }

\hline

Parameter &  TRES & HIRES & TRES and HIRES\\

\hline

$P$ (days) & $88.06$ & $88.193$ & $88.188$ \\

$t_{\rm{tran}}$ (BJD $- \, 2,450,000$) &  $5002.0$  & $4997.2$ & $4997.25$ \\

$e \cos\omega$ &   $0.0$  & $0.0$  & $0.0048$ \\

$e \sin\omega$ &  $ 0.0$ & $0.0$ & $-0.0063$ \\

$K_1$ ($\rm{km \: s^{-1}}$) &   $19.56$ & $19.81$ & $19.77$ \\

$\gamma$ ($\rm{km \: s^{-1}}$) &  $-27.35$ & $-27.3$ & $-27.361$ \\

$\gamma_o$ ($\rm{km \: s^{-1}}$) &  ... & ... & $-0.04$ \\

$\sigma_{\rm{j, HIRES}}$ ($\rm{km \: s^{-1}}$) & ... & $0.045$  & $0.032$ \\

$\sigma_{\rm{j, TRES}}$ ($\rm{km \: s^{-1}}$) &  $0.17$  & ... & $0.202$ \\

$e$  & $0.0$  & $0.0$  & $0.0079$  \\

$\omega$ (deg)  & $-87.0$ & $-56.0$ & $-53.0$   \\

\hline
\end{tabular}
\end{table}

\begin{table}[h]
\centering
\caption{Maximum-Likelihood Parameters from the Dynamical MCMC Model} 
\label{tab: Best-compare_RVmass_results}
\begin{tabular}{ c  c  c  c }
\hline
Parameter & SPC & Brewer & SpecMatch \\
\hline

$M_1$ ($M\textsubscript{\(\odot\)}$) &   $0.891$ & $0.868$ & $0.906$  \\

[Fe/H] &  $0.153$  &  $0.106$ &  $0.162$\\

\hline

\end{tabular}
\end{table}

\begin{table}[h]
\centering
\caption{Maximum-Likelihood Parameters from the Joint Microlensing and Radial Velocity Model}
\label{tab: Best-Joint_model_results}
\begin{tabular}{ c  c  c  c }
\hline
Parameter & SPC & Brewer & SpecMatch \\
\hline 

$P$ (days) & $88.18059$ & $88.18066$  & $88.18048$ \\

$t_{tran}$ (BJD $- \, 2,455,000$) & $85.417$ & $85.4191$  & $85.4194$  \\

$e \cos\omega$  & $0.014753$ & $0.014708$ &  $0.014717$  \\

$e \sin\omega$  & $-0.0112$ & $-0.0092$ &  $-0.0092$  \\

$b$  & $0.643$ &  $0.609$ & $0.656$  \\

$R_2$ ($R\textsubscript{\(\odot\)}$) & $0.0127$ & $0.0143$ & $0.0121$ \\

$M_1$ ($M\textsubscript{\(\odot\)}$) & $0.955$ & $0.923$ & $0.974$ \\

[Fe/H]  & $0.184$ & $0.085$ & $0.149$   \\

Age (Gyr) & $0.9$ & $0.9$  & $0.9$ \\

$K_1$ ($\rm{km \: s^{-1}}$) &   $19.711$ & $19.761$ & $19.684$ \\

$\gamma$ ($\rm{km \: s^{-1}}$) & $-27.461$ & $-27.466$ & $-27.436$ \\

$\gamma_o$ ($\rm{km \: s^{-1}}$) &  $-0.07$ & $-0.07$ & $-0.05$ \\

$\sigma_{\rm{j, HIRES}}$ ($\rm{km \: s^{-1}}$) & $0.032$ & $0.032$  & $0.055$ \\

$\sigma_{\rm{j, TRES}}$ ($\rm{km \: s^{-1}}$) &  $0.221$  & $0.253$ & $0.230$ \\

$F_2/F_1$ & $0.001117$ & $0.001109$ & $0.00114$ \\

$e$  & $0.0185$ & $0.0173$ & $0.0173$ \\

$\omega$ (deg)  & $-37.0$ & $-32.0$ & $-32.0$  \\
\hline
\end{tabular}
\end{table}



\end{document}